\documentclass[longauth]{aa}

\usepackage{graphicx}
\usepackage{natbib}
\usepackage{scalerel}
\usepackage{comment}

\usepackage[table]{xcolor}

\usepackage{txfonts}
\usepackage[pdfencoding=auto,psdextra]{hyperref}
\hypersetup{
    colorlinks=true,
    linkcolor=blue,
    filecolor=magenta,
    urlcolor=blue,
    citecolor=blue
}
\makeatletter
\renewcommand*\aa@pageof{, page \thepage{} of \pageref*{LastPage}}
\makeatother

\usepackage[utf8]{inputenc}

\usepackage[switch, modulo]{lineno}
\renewcommand{\linenumbers}[0]{}
\usepackage{euclid}

\newcommand{\Ha}{\ha}
\newcommand{\Hb}{\hb}

\begin{document}
%
%

\title{\Euclid: Data Release 1 (DR1) -- First NISP emission-line maps:
Methodology, validation, and applications to galaxy evolution\thanks{This paper is published on behalf of the Euclid Consortium}}

\newcommand{\orcid}[1]{} 
\author{L.~Gabarra\orcid{0000-0002-8486-8856}\thanks{\email{louis.gabarra@physics.ox.ac.uk}}\inst{\ref{aff1}}
\and B.~R.~Granett\orcid{0000-0003-2694-9284}\inst{\ref{aff2}}
\and C.~Mancini\orcid{0000-0002-4297-0561}\inst{\ref{aff3}}
\and G.~Dalton\orcid{0000-0002-3031-2588}\inst{\ref{aff1},\ref{aff4}}
\and A.~Henderson~de~la~Fuente\orcid{0009-0007-6670-7657}\inst{\ref{aff1}}
\and F.~Rigamonti\orcid{0000-0001-7880-8825}\inst{\ref{aff5}}
\and S.~Quai\orcid{0000-0002-0449-8163}\inst{\ref{aff6},\ref{aff7}}
\and F.~Ricci\orcid{0000-0001-5742-5980}\inst{\ref{aff8},\ref{aff9}}
\and E.~Daddi\orcid{0000-0002-3331-9590}\inst{\ref{aff10}}
\and M.~Talia\orcid{0000-0003-4352-2063}\inst{\ref{aff6},\ref{aff7}}
\and M.~Fumagalli\orcid{0000-0001-6676-3842}\inst{\ref{aff11},\ref{aff12}}
\and M.~Bolzonella\orcid{0000-0003-3278-4607}\inst{\ref{aff7}}
\and A.~Feltre\orcid{0000-0001-6865-2871}\inst{\ref{aff13}}
\and J.~Matharu\orcid{0000-0002-7547-3385}\inst{\ref{aff14}}
\and T.~Harvey\orcid{0000-0002-4130-636X}\inst{\ref{aff14},\ref{aff15}}
\and D.~Scott\orcid{0000-0002-6878-9840}\inst{\ref{aff16}}
\and G.~Zamorani\orcid{0000-0002-2318-301X}\inst{\ref{aff7}}
\and L.~Pozzetti\orcid{0000-0001-7085-0412}\inst{\ref{aff7}}
\and P.~Cassata\orcid{0000-0002-6716-4400}\inst{\ref{aff17},\ref{aff18}}
\and G.~Rodighiero\orcid{0000-0002-9415-2296}\inst{\ref{aff17},\ref{aff18}}
\and F.~Cogato\orcid{0000-0003-4632-6113}\inst{\ref{aff7}}
\and E.~Medinaceli\orcid{0000-0002-4040-7783}\inst{\ref{aff7}}
\and I.~Lewis\orcid{0009-0009-2090-5513}\inst{\ref{aff1}}
\and E.~Schallig\orcid{0000-0003-2517-0965}\inst{\ref{aff1}}
\and R.~G.~Varadaraj\orcid{0009-0006-9953-6471}\inst{\ref{aff1}}
\and A.~Iovino\orcid{0000-0001-6958-0304}\inst{\ref{aff5}}
\and A.~Mercurio\orcid{0000-0001-9261-7849}\inst{\ref{aff19},\ref{aff20},\ref{aff21}}
\and M.~Longhetti\inst{\ref{aff2}}
\and F.~Shankar\orcid{0000-0001-8973-5051}\inst{\ref{aff22}}
\and B.~Altieri\orcid{0000-0003-3936-0284}\inst{\ref{aff23}}
\and S.~Andreon\orcid{0000-0002-2041-8784}\inst{\ref{aff2}}
\and N.~Auricchio\orcid{0000-0003-4444-8651}\inst{\ref{aff7}}
\and C.~Baccigalupi\orcid{0000-0002-8211-1630}\inst{\ref{aff24},\ref{aff12},\ref{aff25},\ref{aff26}}
\and M.~Baldi\orcid{0000-0003-4145-1943}\inst{\ref{aff27},\ref{aff7},\ref{aff28}}
\and A.~Balestra\orcid{0000-0002-6967-261X}\inst{\ref{aff18}}
\and S.~Bardelli\orcid{0000-0002-8900-0298}\inst{\ref{aff7}}
\and P.~Battaglia\orcid{0000-0002-7337-5909}\inst{\ref{aff7}}
\and A.~Biviano\orcid{0000-0002-0857-0732}\inst{\ref{aff12},\ref{aff24}}
\and E.~Branchini\orcid{0000-0002-0808-6908}\inst{\ref{aff29},\ref{aff30},\ref{aff2}}
\and M.~Brescia\orcid{0000-0001-9506-5680}\inst{\ref{aff31},\ref{aff19}}
\and J.~Brinchmann\orcid{0000-0003-4359-8797}\inst{\ref{aff32},\ref{aff33},\ref{aff34}}
\and S.~Camera\orcid{0000-0003-3399-3574}\inst{\ref{aff35},\ref{aff36},\ref{aff37}}
\and V.~Capobianco\orcid{0000-0002-3309-7692}\inst{\ref{aff37}}
\and C.~Carbone\orcid{0000-0003-0125-3563}\inst{\ref{aff3}}
\and J.~Carretero\orcid{0000-0002-3130-0204}\inst{\ref{aff38},\ref{aff39}}
\and M.~Castellano\orcid{0000-0001-9875-8263}\inst{\ref{aff9}}
\and G.~Castignani\orcid{0000-0001-6831-0687}\inst{\ref{aff7}}
\and S.~Cavuoti\orcid{0000-0002-3787-4196}\inst{\ref{aff19},\ref{aff40}}
\and P.-Y.~Chabaud\orcid{0009-0002-7741-4722}\inst{\ref{aff41}}
\and K.~C.~Chambers\orcid{0000-0001-6965-7789}\inst{\ref{aff42}}
\and A.~Cimatti\inst{\ref{aff43}}
\and C.~Colodro-Conde\inst{\ref{aff44}}
\and G.~Congedo\orcid{0000-0003-2508-0046}\inst{\ref{aff45}}
\and C.~J.~Conselice\orcid{0000-0003-1949-7638}\inst{\ref{aff15}}
\and L.~Conversi\orcid{0000-0002-6710-8476}\inst{\ref{aff46},\ref{aff23}}
\and Y.~Copin\orcid{0000-0002-5317-7518}\inst{\ref{aff47}}
\and A.~Costille\inst{\ref{aff41}}
\and F.~Courbin\orcid{0000-0003-0758-6510}\inst{\ref{aff48},\ref{aff49},\ref{aff50}}
\and H.~M.~Courtois\orcid{0000-0003-0509-1776}\inst{\ref{aff51}}
\and M.~Cropper\orcid{0000-0003-4571-9468}\inst{\ref{aff52}}
\and H.~Degaudenzi\orcid{0000-0002-5887-6799}\inst{\ref{aff53}}
\and S.~de~la~Torre\inst{\ref{aff41}}
\and G.~De~Lucia\orcid{0000-0002-6220-9104}\inst{\ref{aff12}}
\and H.~Dole\orcid{0000-0002-9767-3839}\inst{\ref{aff54},\ref{aff55}}
\and F.~Dubath\orcid{0000-0002-6533-2810}\inst{\ref{aff53}}
\and X.~Dupac\inst{\ref{aff23}}
\and S.~Escoffier\orcid{0000-0002-2847-7498}\inst{\ref{aff56}}
\and M.~Fabricius\orcid{0000-0002-7025-6058}\inst{\ref{aff57},\ref{aff58}}
\and M.~Farina\orcid{0000-0002-3089-7846}\inst{\ref{aff59}}
\and R.~Farinelli\inst{\ref{aff7}}
\and F.~Faustini\orcid{0000-0001-6274-5145}\inst{\ref{aff9}}
\and S.~Ferriol\inst{\ref{aff47}}
\and F.~Finelli\orcid{0000-0002-6694-3269}\inst{\ref{aff7},\ref{aff60}}
\and P.~Fosalba\orcid{0000-0002-1510-5214}\inst{\ref{aff61},\ref{aff62}}
\and N.~Fourmanoit\orcid{0009-0005-6816-6925}\inst{\ref{aff56}}
\and M.~Frailis\orcid{0000-0002-7400-2135}\inst{\ref{aff12}}
\and E.~Franceschi\orcid{0000-0002-0585-6591}\inst{\ref{aff7}}
\and M.~Fumana\orcid{0000-0001-6787-5950}\inst{\ref{aff3}}
\and S.~Galeotta\orcid{0000-0002-3748-5115}\inst{\ref{aff12}}
\and K.~George\orcid{0000-0002-1734-8455}\inst{\ref{aff63}}
\and W.~Gillard\orcid{0000-0003-4744-9748}\inst{\ref{aff56}}
\and C.~Giocoli\orcid{0000-0002-9590-7961}\inst{\ref{aff7},\ref{aff28}}
\and P.~G\'omez-Alvarez\orcid{0000-0002-8594-5358}\inst{\ref{aff64},\ref{aff23}}
\and J.~Gracia-Carpio\orcid{0000-0003-4689-3134}\inst{\ref{aff57}}
\and A.~Grazian\orcid{0000-0002-5688-0663}\inst{\ref{aff18}}
\and F.~Grupp\inst{\ref{aff57},\ref{aff58}}
\and L.~Guzzo\orcid{0000-0001-8264-5192}\inst{\ref{aff65},\ref{aff2},\ref{aff66}}
\and W.~G.~Hartley\orcid{0000-0002-4563-5101}\inst{\ref{aff53}}
\and S.~V.~H.~Haugan\orcid{0000-0001-9648-7260}\inst{\ref{aff67}}
\and S.~Hemmati\orcid{0000-0003-2226-5395}\inst{\ref{aff68}}
\and H.~Hoekstra\orcid{0000-0002-0641-3231}\inst{\ref{aff69}}
\and W.~Holmes\orcid{0009-0007-8554-4646}\inst{\ref{aff70}}
\and I.~M.~Hook\orcid{0000-0002-2960-978X}\inst{\ref{aff71}}
\and F.~Hormuth\inst{\ref{aff72}}
\and A.~Hornstrup\orcid{0000-0002-3363-0936}\inst{\ref{aff73},\ref{aff74}}
\and M.~Huertas-Company\orcid{0000-0002-1416-8483}\inst{\ref{aff44},\ref{aff75},\ref{aff76}}
\and K.~Jahnke\orcid{0000-0003-3804-2137}\inst{\ref{aff14}}
\and M.~Jhabvala\inst{\ref{aff77}}
\and S.~Kermiche\orcid{0000-0002-0302-5735}\inst{\ref{aff56}}
\and A.~Kiessling\orcid{0000-0002-2590-1273}\inst{\ref{aff70}}
\and B.~Kubik\orcid{0009-0006-5823-4880}\inst{\ref{aff47}}
\and M.~K\"ummel\orcid{0000-0003-2791-2117}\inst{\ref{aff58}}
\and M.~Kunz\orcid{0000-0002-3052-7394}\inst{\ref{aff78}}
\and H.~Kurki-Suonio\orcid{0000-0002-4618-3063}\inst{\ref{aff79},\ref{aff80}}
\and A.~M.~C.~Le~Brun\orcid{0000-0002-0936-4594}\inst{\ref{aff81}}
\and V.~Le~Brun\orcid{0000-0002-5027-1939}\inst{\ref{aff41}}
\and D.~Le~Mignant\orcid{0000-0002-5339-5515}\inst{\ref{aff41}}
\and P.~Liebing\orcid{0009-0005-8043-7150}\inst{\ref{aff52}}
\and S.~Ligori\orcid{0000-0003-4172-4606}\inst{\ref{aff37}}
\and P.~B.~Lilje\orcid{0000-0003-4324-7794}\inst{\ref{aff67}}
\and V.~Lindholm\orcid{0000-0003-2317-5471}\inst{\ref{aff79},\ref{aff80}}
\and I.~Lloro\orcid{0000-0001-5966-1434}\inst{\ref{aff82}}
\and M.~Magliocchetti\orcid{0000-0001-9158-4838}\inst{\ref{aff59}}
\and G.~Mainetti\orcid{0000-0003-2384-2377}\inst{\ref{aff83}}
\and O.~Mansutti\orcid{0000-0001-5758-4658}\inst{\ref{aff12}}
\and S.~Marcin\inst{\ref{aff84}}
\and O.~Marggraf\orcid{0000-0001-7242-3852}\inst{\ref{aff85}}
\and M.~Martinelli\orcid{0000-0002-6943-7732}\inst{\ref{aff9},\ref{aff86}}
\and N.~Martinet\orcid{0000-0003-2786-7790}\inst{\ref{aff41}}
\and F.~Marulli\orcid{0000-0002-8850-0303}\inst{\ref{aff6},\ref{aff7},\ref{aff28}}
\and R.~J.~Massey\orcid{0000-0002-6085-3780}\inst{\ref{aff87}}
\and S.~Mei\orcid{0000-0002-2849-559X}\inst{\ref{aff88},\ref{aff89}}
\and M.~Meneghetti\orcid{0000-0003-1225-7084}\inst{\ref{aff7},\ref{aff28}}
\and E.~Merlin\orcid{0000-0001-6870-8900}\inst{\ref{aff18}}
\and G.~Meylan\orcid{0000-0001-6503-0209}\inst{\ref{aff90}}
\and P.~Monaco\orcid{0000-0003-2083-7564}\inst{\ref{aff91},\ref{aff12},\ref{aff25},\ref{aff24}}
\and A.~Mora\orcid{0000-0002-1922-8529}\inst{\ref{aff92}}
\and M.~Moresco\orcid{0000-0002-7616-7136}\inst{\ref{aff6},\ref{aff7}}
\and C.~Moretti\orcid{0000-0003-3314-8936}\inst{\ref{aff12},\ref{aff24},\ref{aff25}}
\and L.~Moscardini\orcid{0000-0002-3473-6716}\inst{\ref{aff6},\ref{aff7},\ref{aff28}}
\and R.~Nakajima\orcid{0009-0009-1213-7040}\inst{\ref{aff85}}
\and C.~Neissner\orcid{0000-0001-8524-4968}\inst{\ref{aff93},\ref{aff39}}
\and S.-M.~Niemi\orcid{0009-0005-0247-0086}\inst{\ref{aff94}}
\and C.~Padilla\orcid{0000-0001-7951-0166}\inst{\ref{aff93}}
\and S.~Paltani\orcid{0000-0002-8108-9179}\inst{\ref{aff53}}
\and F.~Pasian\orcid{0000-0002-4869-3227}\inst{\ref{aff12}}
\and W.~J.~Percival\orcid{0000-0002-0644-5727}\inst{\ref{aff95},\ref{aff96},\ref{aff97}}
\and V.~Pettorino\orcid{0000-0002-4203-9320}\inst{\ref{aff94}}
\and A.~Pezzotta\orcid{0000-0003-0726-2268}\inst{\ref{aff2}}
\and G.~Polenta\orcid{0000-0003-4067-9196}\inst{\ref{aff98}}
\and M.~Poncet\inst{\ref{aff99}}
\and L.~A.~Popa\inst{\ref{aff100}}
\and F.~Raison\orcid{0000-0002-7819-6918}\inst{\ref{aff57}}
\and A.~Renzi\orcid{0000-0001-9856-1970}\inst{\ref{aff17},\ref{aff101},\ref{aff7}}
\and J.~Rhodes\orcid{0000-0002-4485-8549}\inst{\ref{aff70}}
\and G.~Riccio\inst{\ref{aff19}}
\and I.~Risso\orcid{0000-0003-2525-7761}\inst{\ref{aff29},\ref{aff30},\ref{aff2}}
\and F.~Rizzo\orcid{0000-0002-9407-585X}\inst{\ref{aff12}}
\and E.~Romelli\orcid{0000-0003-3069-9222}\inst{\ref{aff12}}
\and M.~Roncarelli\orcid{0000-0001-9587-7822}\inst{\ref{aff7}}
\and B.~Rusholme\orcid{0000-0001-7648-4142}\inst{\ref{aff68}}
\and R.~Saglia\orcid{0000-0003-0378-7032}\inst{\ref{aff58},\ref{aff57}}
\and Z.~Sakr\orcid{0000-0002-4823-3757}\inst{\ref{aff102},\ref{aff103},\ref{aff104}}
\and D.~Sapone\orcid{0000-0001-7089-4503}\inst{\ref{aff105}}
\and B.~Sartoris\orcid{0000-0003-1337-5269}\inst{\ref{aff106}}
\and M.~Sauvage\orcid{0000-0002-0809-2574}\inst{\ref{aff10}}
\and M.~Schirmer\orcid{0000-0003-2568-9994}\inst{\ref{aff14}}
\and P.~Schneider\orcid{0000-0001-8561-2679}\inst{\ref{aff85}}
\and M.~Scodeggio\inst{\ref{aff3}}
\and A.~Secroun\orcid{0000-0003-0505-3710}\inst{\ref{aff56}}
\and E.~Sihvola\orcid{0000-0003-1804-7715}\inst{\ref{aff79},\ref{aff80}}
\and P.~Simon\inst{\ref{aff85}}
\and C.~Sirignano\orcid{0000-0002-0995-7146}\inst{\ref{aff17},\ref{aff101}}
\and G.~Sirri\orcid{0000-0003-2626-2853}\inst{\ref{aff28}}
\and L.~Stanco\orcid{0000-0002-9706-5104}\inst{\ref{aff101}}
\and C.~Surace\orcid{0000-0003-2592-0113}\inst{\ref{aff41}}
\and P.~Tallada-Cresp\'{i}\orcid{0000-0002-1336-8328}\inst{\ref{aff38},\ref{aff39}}
\and A.~N.~Taylor\inst{\ref{aff45}}
\and H.~I.~Teplitz\orcid{0000-0002-7064-5424}\inst{\ref{aff107}}
\and I.~Tereno\orcid{0000-0002-4537-6218}\inst{\ref{aff108},\ref{aff109}}
\and S.~Toft\orcid{0000-0003-3631-7176}\inst{\ref{aff110},\ref{aff111}}
\and R.~Toledo-Moreo\orcid{0000-0002-2997-4859}\inst{\ref{aff112},\ref{aff113}}
\and F.~Torradeflot\orcid{0000-0003-1160-1517}\inst{\ref{aff39},\ref{aff38}}
\and I.~Tutusaus\orcid{0000-0002-3199-0399}\inst{\ref{aff62},\ref{aff61},\ref{aff103}}
\and J.~Valiviita\orcid{0000-0001-6225-3693}\inst{\ref{aff79},\ref{aff80}}
\and T.~Vassallo\orcid{0000-0001-6512-6358}\inst{\ref{aff12},\ref{aff63}}
\and D.~Vibert\orcid{0009-0008-0607-631X}\inst{\ref{aff41}}
\and Y.~Wang\orcid{0000-0002-4749-2984}\inst{\ref{aff68}}
\and J.~Weller\orcid{0000-0002-8282-2010}\inst{\ref{aff58},\ref{aff57}}
\and A.~Zacchei\orcid{0000-0003-0396-1192}\inst{\ref{aff12},\ref{aff24}}
\and F.~M.~Zerbi\orcid{0000-0002-9996-973X}\inst{\ref{aff2}}
\and I.~A.~Zinchenko\orcid{0000-0002-2944-2449}\inst{\ref{aff114}}
\and E.~Zucca\orcid{0000-0002-5845-8132}\inst{\ref{aff7}}
\and P.~Bergamini\orcid{0000-0003-1383-9414}\inst{\ref{aff7}}
\and J.~Garc\'ia-Bellido\orcid{0000-0002-9370-8360}\inst{\ref{aff102}}
\and C.~M.~Gutierrez\orcid{0000-0001-7854-783X}\inst{\ref{aff44},\ref{aff115}}
\and K.~Tanidis\orcid{0000-0001-9843-5130}\inst{\ref{aff116}}}
										   
\institute{Department of Physics, University of Oxford, Keble Road, Oxford OX1 3RH, UK\label{aff1}
\and
INAF-Osservatorio Astronomico di Brera, Via Brera 28, 20122 Milano, Italy\label{aff2}
\and
INAF-IASF Milano, Via Alfonso Corti 12, 20133 Milano, Italy\label{aff3}
\and
RAL Space, Rutherford Appleton Laboratory, STFC, UKRI, Harwell Campus, Oxfordshire, OX11 0QX, UK\label{aff4}
\and
INAF - Osservatorio Astronomico di Brera, via Emilio Bianchi 46, 23807 Merate, Italy\label{aff5}
\and
Dipartimento di Fisica e Astronomia "Augusto Righi" - Alma Mater Studiorum Universit\`a di Bologna, via Piero Gobetti 93/2, 40129 Bologna, Italy\label{aff6}
\and
INAF-Osservatorio di Astrofisica e Scienza dello Spazio di Bologna, Via Piero Gobetti 93/3, 40129 Bologna, Italy\label{aff7}
\and
Department of Mathematics and Physics, Roma Tre University, Via della Vasca Navale 84, 00146 Rome, Italy\label{aff8}
\and
INAF-Osservatorio Astronomico di Roma, Via Frascati 33, 00078 Monteporzio Catone, Italy\label{aff9}
\and
Universit\'e Paris-Saclay, Universit\'e Paris Cit\'e, CEA, CNRS, AIM, 91191, Gif-sur-Yvette, France\label{aff10}
\and
Dipartimento di Fisica ``G. Occhialini", Universit\`a degli Studi di Milano Bicocca, Piazza della Scienza 3, 20126 Milano, Italy\label{aff11}
\and
INAF-Osservatorio Astronomico di Trieste, Via G. B. Tiepolo 11, 34143 Trieste, Italy\label{aff12}
\and
INAF-Osservatorio Astrofisico di Arcetri, Largo E. Fermi 5, 50125, Firenze, Italy\label{aff13}
\and
Max-Planck-Institut f\"ur Astronomie, K\"onigstuhl 17, 69117 Heidelberg, Germany\label{aff14}
\and
Jodrell Bank Centre for Astrophysics, Department of Physics and Astronomy, University of Manchester, Oxford Road, Manchester M13 9PL, UK\label{aff15}
\and
Department of Physics and Astronomy, University of British Columbia, Vancouver, BC V6T 1Z1, Canada\label{aff16}
\and
Dipartimento di Fisica e Astronomia "G. Galilei", Universit\`a di Padova, Via Marzolo 8, 35131 Padova, Italy\label{aff17}
\and
INAF-Osservatorio Astronomico di Padova, Via dell'Osservatorio 5, 35122 Padova, Italy\label{aff18}
\and
INAF-Osservatorio Astronomico di Capodimonte, Via Moiariello 16, 80131 Napoli, Italy\label{aff19}
\and
Universita di Salerno, Dipartimento di Fisica "E.R. Caianiello", Via Giovanni Paolo II 132, I-84084 Fisciano (SA), Italy\label{aff20}
\and
INFN -- Gruppo Collegato di Salerno - Sezione di Napoli, Dipartimento di Fisica "E.R. Caianiello", Universita di Salerno, via Giovanni Paolo II, 132 - I-84084 Fisciano (SA), Italy\label{aff21}
\and
School of Physics \& Astronomy, University of Southampton, Highfield Campus, Southampton SO17 1BJ, UK\label{aff22}
\and
ESAC/ESA, Camino Bajo del Castillo, s/n., Urb. Villafranca del Castillo, 28692 Villanueva de la Ca\~nada, Madrid, Spain\label{aff23}
\and
IFPU, Institute for Fundamental Physics of the Universe, via Beirut 2, 34151 Trieste, Italy\label{aff24}
\and
INFN, Sezione di Trieste, Via Valerio 2, 34127 Trieste TS, Italy\label{aff25}
\and
SISSA, International School for Advanced Studies, Via Bonomea 265, 34136 Trieste TS, Italy\label{aff26}
\and
Dipartimento di Fisica e Astronomia, Universit\`a di Bologna, Via Gobetti 93/2, 40129 Bologna, Italy\label{aff27}
\and
INFN-Sezione di Bologna, Viale Berti Pichat 6/2, 40127 Bologna, Italy\label{aff28}
\and
Dipartimento di Fisica, Universit\`a di Genova, Via Dodecaneso 33, 16146, Genova, Italy\label{aff29}
\and
INFN-Sezione di Genova, Via Dodecaneso 33, 16146, Genova, Italy\label{aff30}
\and
Department of Physics "E. Pancini", University Federico II, Via Cinthia 6, 80126, Napoli, Italy\label{aff31}
\and
European Southern Observatory, Karl-Schwarzschild-Str.~2, 85748 Garching, Germany\label{aff32}
\and
Instituto de Astrof\'isica e Ci\^encias do Espa\c{c}o, Universidade do Porto, CAUP, Rua das Estrelas, PT4150-762 Porto, Portugal\label{aff33}
\and
Faculdade de Ci\^encias da Universidade do Porto, Rua do Campo de Alegre, 4150-007 Porto, Portugal\label{aff34}
\and
Dipartimento di Fisica, Universit\`a degli Studi di Torino, Via P. Giuria 1, 10125 Torino, Italy\label{aff35}
\and
INFN-Sezione di Torino, Via P. Giuria 1, 10125 Torino, Italy\label{aff36}
\and
INAF-Osservatorio Astrofisico di Torino, Via Osservatorio 20, 10025 Pino Torinese (TO), Italy\label{aff37}
\and
Centro de Investigaciones Energ\'eticas, Medioambientales y Tecnol\'ogicas (CIEMAT), Avenida Complutense 40, 28040 Madrid, Spain\label{aff38}
\and
Port d'Informaci\'{o} Cient\'{i}fica, Campus UAB, C. Albareda s/n, 08193 Bellaterra (Barcelona), Spain\label{aff39}
\and
INFN -- Sezione di Napoli, Via Cinthia 6, 80126, Napoli, Italy\label{aff40}
\and
Aix-Marseille Universit\'e, CNRS, CNES, LAM, Marseille, France\label{aff41}
\and
Institute for Astronomy, University of Hawaii, 2680 Woodlawn Drive, Honolulu, HI 96822, USA\label{aff42}
\and
Dipartimento di Fisica e Astronomia "Augusto Righi" - Alma Mater Studiorum Universit\`a di Bologna, Viale Berti Pichat 6/2, 40127 Bologna, Italy\label{aff43}
\and
Instituto de Astrof\'{\i}sica de Canarias, E-38205 La Laguna, Tenerife, Spain\label{aff44}
\and
Institute for Astronomy, University of Edinburgh, Royal Observatory, Blackford Hill, Edinburgh EH9 3HJ, UK\label{aff45}
\and
European Space Agency/ESRIN, Largo Galileo Galilei 1, 00044 Frascati, Roma, Italy\label{aff46}
\and
Universit\'e Lyon 1, CNRS/IN2P3, IP2I Lyon, Villeurbanne, France\label{aff47}
\and
Institut de Ci\`{e}ncies del Cosmos (ICCUB), Universitat de Barcelona (IEEC-UB), Mart\'{i} i Franqu\`{e}s 1, 08028 Barcelona, Spain\label{aff48}
\and
Instituci\'o Catalana de Recerca i Estudis Avan\c{c}ats (ICREA), Passeig de Llu\'{\i}s Companys 23, 08010 Barcelona, Spain\label{aff49}
\and
Institut de Ciencies de l'Espai (IEEC-CSIC), Campus UAB, Carrer de Can Magrans, s/n Cerdanyola del Vall\'es, 08193 Barcelona, Spain\label{aff50}
\and
Universit\'e Lyon 1, CNRS/IN2P3, IUF, IP2I Lyon, Villeurbanne, France \label{aff51}
\and
Mullard Space Science Laboratory, University College London, Holmbury St Mary, Dorking, Surrey RH5 6NT, UK\label{aff52}
\and
Department of Astronomy, University of Geneva, ch. d'Ecogia 16, 1290 Versoix, Switzerland\label{aff53}
\and
Universit\'e Paris-Saclay, CNRS, Institut d'astrophysique spatiale, 91405, Orsay, France\label{aff54}
\and
Institut universitaire de France (IUF), 1 rue Descartes, 75231 PARIS CEDEX 05, France\label{aff55}
\and
Aix-Marseille Universit\'e, CNRS/IN2P3, CPPM, Marseille, France\label{aff56}
\and
Max Planck Institute for Extraterrestrial Physics, Giessenbachstr. 1, 85748 Garching, Germany\label{aff57}
\and
Universit\"ats-Sternwarte M\"unchen, Fakult\"at f\"ur Physik, Ludwig-Maximilians-Universit\"at M\"unchen, Scheinerstr.~1, 81679 M\"unchen, Germany\label{aff58}
\and
INAF-Istituto di Astrofisica e Planetologia Spaziali, via del Fosso del Cavaliere, 100, 00100 Roma, Italy\label{aff59}
\and
INFN-Bologna, Via Irnerio 46, 40126 Bologna, Italy\label{aff60}
\and
Institut d'Estudis Espacials de Catalunya (IEEC),  Edifici RDIT, Campus UPC, 08860 Castelldefels, Barcelona, Spain\label{aff61}
\and
Institute of Space Sciences (ICE, CSIC), Campus UAB, Carrer de Can Magrans, s/n, 08193 Barcelona, Spain\label{aff62}
\and
University Observatory, LMU Faculty of Physics, Scheinerstr.~1, 81679 Munich, Germany\label{aff63}
\and
FRACTAL S.L.N.E., calle Tulip\'an 2, Portal 13 1A, 28231, Las Rozas de Madrid, Spain\label{aff64}
\and
Dipartimento di Fisica "Aldo Pontremoli", Universit\`a degli Studi di Milano, Via Celoria 16, 20133 Milano, Italy\label{aff65}
\and
INFN-Sezione di Milano, Via Celoria 16, 20133 Milano, Italy\label{aff66}
\and
Institute of Theoretical Astrophysics, University of Oslo, P.O. Box 1029 Blindern, 0315 Oslo, Norway\label{aff67}
\and
Caltech/IPAC, 1200 E. California Blvd., Pasadena, CA 91125, USA\label{aff68}
\and
Leiden Observatory, Leiden University, Einsteinweg 55, 2333 CC Leiden, The Netherlands\label{aff69}
\and
Jet Propulsion Laboratory, California Institute of Technology, 4800 Oak Grove Drive, Pasadena, CA, 91109, USA\label{aff70}
\and
School of Physics and Astronomy, Lancaster University, Lancaster, LA1 4YB, UK\label{aff71}
\and
Felix Hormuth Engineering, Goethestr. 17, 69181 Leimen, Germany\label{aff72}
\and
Technical University of Denmark, Elektrovej 327, 2800 Kgs. Lyngby, Denmark\label{aff73}
\and
Cosmic Dawn Center (DAWN), Denmark\label{aff74}
\and
Universit\'e PSL, Observatoire de Paris, Sorbonne Universit\'e, CNRS, LERMA, 75014, Paris, France\label{aff75}
\and
Universit\'e Paris-Cit\'e, 5 Rue Thomas Mann, 75013, Paris, France\label{aff76}
\and
NASA Goddard Space Flight Center, Greenbelt, MD 20771, USA\label{aff77}
\and
Universit\'e de Gen\`eve, D\'epartement de Physique Th\'eorique and Centre for Astroparticle Physics, 24 quai Ernest-Ansermet, CH-1211 Gen\`eve 4, Switzerland\label{aff78}
\and
Department of Physics, P.O. Box 64, University of Helsinki, 00014 Helsinki, Finland\label{aff79}
\and
Helsinki Institute of Physics, Gustaf H{\"a}llstr{\"o}min katu 2, University of Helsinki, 00014 Helsinki, Finland\label{aff80}
\and
Laboratoire d'etude de l'Univers et des phenomenes eXtremes, Observatoire de Paris, Universit\'e PSL, Sorbonne Universit\'e, CNRS, 92190 Meudon, France\label{aff81}
\and
SKAO, Jodrell Bank, Lower Withington, Macclesfield SK11 9FT, UK\label{aff82}
\and
Centre de Calcul de l'IN2P3/CNRS, 21 avenue Pierre de Coubertin 69627 Villeurbanne Cedex, France\label{aff83}
\and
University of Applied Sciences and Arts of Northwestern Switzerland, School of Computer Science, 5210 Windisch, Switzerland\label{aff84}
\and
Universit\"at Bonn, Argelander-Institut f\"ur Astronomie, Auf dem H\"ugel 71, 53121 Bonn, Germany\label{aff85}
\and
INFN-Sezione di Roma, Piazzale Aldo Moro, 2 - c/o Dipartimento di Fisica, Edificio G. Marconi, 00185 Roma, Italy\label{aff86}
\and
Department of Physics, Institute for Computational Cosmology, Durham University, South Road, Durham, DH1 3LE, UK\label{aff87}
\and
Universit\'e Paris Cit\'e, CNRS, Astroparticule et Cosmologie, 75013 Paris, France\label{aff88}
\and
CNRS-UCB International Research Laboratory, Centre Pierre Bin\'etruy, IRL2007, CPB-IN2P3, Berkeley, USA\label{aff89}
\and
Institute of Physics, Laboratory of Astrophysics, Ecole Polytechnique F\'ed\'erale de Lausanne (EPFL), Observatoire de Sauverny, 1290 Versoix, Switzerland\label{aff90}
\and
Dipartimento di Fisica - Sezione di Astronomia, Universit\`a di Trieste, Via Tiepolo 11, 34131 Trieste, Italy\label{aff91}
\and
Telespazio UK S.L. for European Space Agency (ESA), Camino bajo del Castillo, s/n, Urbanizacion Villafranca del Castillo, Villanueva de la Ca\~nada, 28692 Madrid, Spain\label{aff92}
\and
Institut de F\'{i}sica d'Altes Energies (IFAE), The Barcelona Institute of Science and Technology, Campus UAB, 08193 Bellaterra (Barcelona), Spain\label{aff93}
\and
European Space Agency/ESTEC, Keplerlaan 1, 2201 AZ Noordwijk, The Netherlands\label{aff94}
\and
Waterloo Centre for Astrophysics, University of Waterloo, Waterloo, Ontario N2L 3G1, Canada\label{aff95}
\and
Department of Physics and Astronomy, University of Waterloo, Waterloo, Ontario N2L 3G1, Canada\label{aff96}
\and
Perimeter Institute for Theoretical Physics, Waterloo, Ontario N2L 2Y5, Canada\label{aff97}
\and
Space Science Data Center, Italian Space Agency, via del Politecnico snc, 00133 Roma, Italy\label{aff98}
\and
Centre National d'Etudes Spatiales -- Centre spatial de Toulouse, 18 avenue Edouard Belin, 31401 Toulouse Cedex 9, France\label{aff99}
\and
Institute of Space Science, Str. Atomistilor, nr. 409 M\u{a}gurele, Ilfov, 077125, Romania\label{aff100}
\and
INFN-Padova, Via Marzolo 8, 35131 Padova, Italy\label{aff101}
\and
Instituto de F\'isica Te\'orica UAM-CSIC, Campus de Cantoblanco, 28049 Madrid, Spain\label{aff102}
\and
Institut de Recherche en Astrophysique et Plan\'etologie (IRAP), Universit\'e de Toulouse, CNRS, UPS, CNES, 14 Av. Edouard Belin, 31400 Toulouse, France\label{aff103}
\and
Universit\'e St Joseph; Faculty of Sciences, Beirut, Lebanon\label{aff104}
\and
Departamento de F\'isica, FCFM, Universidad de Chile, Blanco Encalada 2008, Santiago, Chile\label{aff105}
\and
Universit\'e C\^{o}te d'Azur, Observatoire de la C\^{o}te d'Azur, CNRS, Laboratoire Lagrange, Bd de l'Observatoire, CS 34229, 06304 Nice cedex 4, France\label{aff106}
\and
Infrared Processing and Analysis Center, California Institute of Technology, Pasadena, CA 91125, USA\label{aff107}
\and
Departamento de F\'isica, Faculdade de Ci\^encias, Universidade de Lisboa, Edif\'icio C8, Campo Grande, PT1749-016 Lisboa, Portugal\label{aff108}
\and
Instituto de Astrof\'isica e Ci\^encias do Espa\c{c}o, Faculdade de Ci\^encias, Universidade de Lisboa, Campo Grande, 1749-016 Lisboa, Portugal\label{aff109}
\and
Cosmic Dawn Center (DAWN)\label{aff110}
\and
Niels Bohr Institute, University of Copenhagen, Jagtvej 128, 2200 Copenhagen, Denmark\label{aff111}
\and
Universidad Polit\'ecnica de Cartagena, Departamento de Electr\'onica y Tecnolog\'ia de Computadoras,  Plaza del Hospital 1, 30202 Cartagena, Spain\label{aff112}
\and
European University of Technology EUt+, European Union\label{aff113}
\and
Astronomisches Rechen-Institut, Zentrum f\"ur Astronomie der Universit\"at Heidelberg, M\"onchhofstr. 12-14, 69120 Heidelberg, Germany\label{aff114}
\and
Universidad de La Laguna, Dpto. Astrof\'\i sica, E-38206 La Laguna, Tenerife, Spain\label{aff115}
\and
Center for Astrophysics and Cosmology, University of Nova Gorica, Nova Gorica, Slovenia\label{aff116}}      


\date{\today}
%
%
\abstract{We demonstrate the capability of the slitless spectroscopic mode of the \Euclid\ Near-Infrared Spectrometer and Photometer (NISP) to produce emission-line maps analogous to narrow-band images with user-defined central wavelengths. We present the first such maps of galaxies at $0.4 < z < 0.9$, constructed from blue-grism exposures obtained during the \Euclid\ performance-verification observations of the Cosmic Evolution Survey (COSMOS) field in November 2023. Source-injection simulations using four dithered blue-grism exposures yield a mean \Ha surface-brightness sensitivity of $2.0\times10^{-16}\,\mathrm{erg\,s^{-1}\,cm^{-2}\,arcsec^{-2}}$ at ${\rm S/N}=7$, averaged within the point-spread function (PSF)-convolved \Ha half-light radius. For sources above this threshold and with \YE>19.9, the median fractional \Ha flux residual remains within approximately $5\%$, with a scaled median absolute deviation of $14\%$ and $24\%$ for intrinsic full widths at half maximum values of \ang{;;0.1} and \ang{;;0.5}, respectively. We measure a median PSF encircled-energy radius of ${\rm EE50}=0\farcs265$, corresponding to a physical radius of $1.4$--$2.3\,\mathrm{kpc}$ over $0.4 < z < 1.8$, the redshift range over which \Ha can be observed with the blue and red grisms. We demonstrate the use of these maps for identifying star-forming clumps, measuring radial star-formation profiles, detecting projected internal velocity gradients, deblending the \Ha--[\ion{N}{ii}] complex, and assessing spectroscopic redshift measurements. As a first statistical application, we reconstruct the galaxy mass--metallicity relation for 885 galaxies observed with the blue grism at $0.4 < z < 0.9$ by deblending the \Ha--[\ion{N}{ii}] complex in the emission-line maps, and measure stacked \Ha radial profiles for 2634 galaxies over the same redshift range. The mass--metallicity relation is consistent with that obtained from the \Euclid 1D spectroscopic measurements, demonstrating overall agreement between the two deblending approaches. The recovered \Ha\ radial profiles broadly follow the \YE\ stellar continuum but remain modestly more extended after matching the \YE\ images to the \Ha\ PSF, consistent with previous results in the literature. Across the four \YE\ size bins, we measure a mean $R_{50,\Ha}/R_{50,\YE}=1.15$.}
%
%
   \titlerunning{\Euclid emission-line maps}
   \authorrunning{L. Gabarra et al.}

   \maketitle

\begin{table*}
\caption{Galaxy targets.}
\label{tab:galaxies}
\centering
\small
\begin{tabular}{l l l l >{\raggedright\arraybackslash}p{6.8cm}}
\hline\hline
\noalign{\vskip 1pt}
Identifier & RA (J2000) & Dec (J2000) & $z$ & Comments \\
\hline
\noalign{\vskip 1pt}
J100106.88+015815.4 & $10^{\rm h}01^{\rm m}06\fs88$ & $+01\degr58\arcmin15\farcs4$ & 0.528 & Fig.~\ref{fig:median_filter}; K26 ID 454417, $z_{\rm K26}=0.5279$ (hCOSMOS) \\
J095955.98+022847.1 & $09^{\rm h}59^{\rm m}55\fs98$ & $+02\degr28\arcmin47\farcs1$ & 0.7043\tablefootmark{a} & Fig.~\ref{fig:face-on}; K26 ID 222529, $z_{\rm K26}=0.7041$ (DESI) \\
J100059.32+022247.7 & $10^{\rm h}00^{\rm m}59\fs32$ & $+02\degr22\arcmin47\farcs7$ & 0.6959 & Fig.~\ref{fig:rotation_vel}; K26 ID 222229, $z_{\rm K26}=0.6966$ (DESI) \\
J100038.66+021643.3 & $10^{\rm h}00^{\rm m}38\fs66$ & $+02\degr16\arcmin43\farcs3$ & 0.4267 & Fig.~\ref{fig:double-dot}; K26 ID 223297, $z_{\rm K26}=0.4265$ (DESI) \\
J095936.59+020647.9 & $09^{\rm h}59^{\rm m}36\fs59$ & $+02\degr06\arcmin47\farcs9$ & 0.5366 & Fig.~\ref{fig:Cluster_and_AGN}a--c; K26 ID 215151, $z_{\rm K26}=0.5360$ (DESI) \\
J100022.52+023014.2 & $10^{\rm h}00^{\rm m}22\fs52$ & $+02\degr30\arcmin14\farcs2$ & 0.751 & Fig.~\ref{fig:Cluster_and_AGN}d--f; K26 ID 385301, $z_{\rm K26}=0.7510$ (HALO7D) \\
J100011.59+023530.1 & $10^{\rm h}00^{\rm m}11\fs59$ & $+02\degr35\arcmin30\farcs1$ & 0.8003\tablefootmark{b} & Fig.~\ref{fig:redshift_refinement}; K26 ID 222990, $z_{\rm K26}=0.8006$ (DESI) \\
J100030.82+022510.0 & $10^{\rm h}00^{\rm m}30\fs82$ & $+02\degr25\arcmin10\farcs0$ & 0.7405\tablefootmark{a} & Fig.~\ref{fig:residual}; K26 ID 223435, $z_{\rm K26}=0.7401$ (DESI)\\
J100004.73+015702.3 & $10^{\rm h}00^{\rm m}04\fs73$ & $+01\degr57\arcmin02\farcs3$ & 0.4390 & Fig.~\ref{fig:smallpacman}; K26 ID 189649, $z_{\rm K26}=0.4386$ (DESI) \\
\hline
\end{tabular}
\tablefoot{The $z$ column gives the adopted spectroscopic redshift,
taken from \citet[K26]{Khostovan2026} unless otherwise indicated.
The Comments column lists the K26 unique-catalogue identifier
(\texttt{Id\_specz}, DR1.1), redshift, and parent survey.
\tablefoottext{a}{SPE redshift.}
\tablefoottext{b}{\texttt{GELSA} redshift estimated from spatial registration; the SPE secondary solution is $z=0.7942$.}}
\end{table*}

\section{\label{sc:NISP}Introduction}

The Near-infrared Spectrometer and Photometer \citep[NISP;][]{EuclidSkyNISP} on board \Euclid\ \citep{EuclidSkyOverview} performs slitless spectroscopy, dispersing the light from every source across the $0.5\,{\rm deg}\textsuperscript{2}$ field of view. Its blue spectroscopic channel covers 926--1366\,nm with a dispersion of 1.24\,nm\,pixel$^{-1}$ and a resolving power of $R=440$--900 for a $0\farcs5$-diameter source. The red channel covers 1206--1892\,nm with a dispersion of 1.37\,nm\,pixel$^{-1}$ and $R=550$--740 \citep{EP-Gillard}. The native detector sampling is $0\farcs3$\,pixel$^{-1}$.

The Euclid Wide Survey \citep[EWS;][]{Scaramella-EP1} will cover \num{14000}\,deg\textsuperscript{2} by the end of the nominal mission using four red-grism exposures obtained at different dispersion orientations ($0^\circ$, $180^\circ$, $-4^\circ$, and $184^\circ)$. The EWS requirement is a $3.5\,\sigma$ emission-line flux sensitivity of $2\times10^{-16}\,\mathrm{erg\,s^{-1}\,cm^{-2}}$ at $1.6\, \micron$ for a \ang{;;0.5}-diameter source, with the first on-sky measurements confirming that NISP meets this requirement \citep{Q1-TP007}. The Euclid Deep Survey (EDS) will cover approximately 50\,deg\textsuperscript{2} through repeated observations with both the red and blue grisms over the nominal 6-year mission \citep{Scaramella-EP1,EuclidSkyOverview}. At the end of the mission, the EDS is expected to reach line-flux sensitivities of $5\times10^{-17}\,\mathrm{erg\,s^{-1}\,cm^{-2}}$ and $6\times10^{-17}\,\mathrm{erg\,s^{-1}\,cm^{-2}}$ with the red and blue grisms, respectively \citep{EuclidSkyOverview}.

The performance of NISP for slitless one-dimensional (1D) spectroscopy on real data is described in \citet{Q1-TP006}, and the effects of galaxy morphology on spectral extraction and emission-line detectability have been investigated using simulations in \citet{Gabarra-EP31}; we also refer to \citet{Walsh2010}, \citet{Momcheva2015}, and \citet{Ryan2018} for further discussion of slitless spectroscopy. Here, we address the complementary problem of reconstructing two-dimensional (2D) emission-line maps from the dispersed NISP observations. In slitless observations, the stellar continuum produces extended spectral traces that may overlap across the field, whereas strong emission lines appear as relatively compact features superimposed on these traces. After subtracting the stellar continuum and mapping the line signal back onto sky coordinates, it is possible to reconstruct a 2D image of the line-emitting regions. Related techniques for recovering spatially resolved emission-line information have been demonstrated with slitless observations from the \HST (HST) and \textit{James Webb} Space Telescope \citep[JWST; for example][]{Nelson_2012,Matharu_2024,Estrada-Carpenter_2024}. In addition, a spectral datacube can be reconstructed from the combination of many slitless spectroscopic exposures with different dispersion orientations \citep{Griggio2026, zhang_2026}.

Spatial information is lost in conventional spectroscopic extractions when a 2D spectrogram is collapsed into a 1D spectrum. For resolved galaxies, spatial and spectral features are superimposed along the dispersion direction, such that emission from distinct regions or nearby spectral features can become confused during the 1D extraction.
The relative orientation of spatial features in a resolved galaxy with respect to the dispersion direction depends on the position angle of the observations and the grism orientation. This further complicates the interpretation when spectrograms are co-added. In contrast, emission-line maps preserve the spatial distribution of the line emission, facilitating its interpretation and enabling spatially resolved studies.

We describe the data set used in this paper in Sect.~\ref{sc:Datasets}, the methodology implemented in the \texttt{GELSA} software\footnote{\url{https://github.com/elsa-euclid/gelsa}} in Sect.~\ref{sc:Gelsa}, performance assessments based on real and simulated data in Sect.~\ref{sc:assessment}, and illustrative scientific applications of \Euclid\ emission-line maps in Sect.~\ref{sc:exampleselmap}. In Sect.~\ref{sect:6}, we demonstrate their potential for statistical studies by reconstructing the galaxy mass--metallicity relation from the \Ha--[\ion{N}{ii}] complex and analysing spatially resolved \Ha radial emission profiles.

In this paper, emission-line wavelengths are specified in vacuum. We adopt a $\Lambda$CDM cosmology with $\Omega_{\mathrm{m}}=0.3$, $\Omega_{\Lambda}=0.7$, and $H_0=70\,\mathrm{km\,s^{-1}\,Mpc^{-1}}$. Clumps in the emission-line maps were identified using the \texttt{DAOStarFinder} implementation available in the \texttt{photutils} Python package\footnote{\url{https://photutils.readthedocs.io/en/stable/}}, adopting a full width at half maximum (FWHM) of 4.5 pixels and a detection threshold of $3\,\sigma$.

\section{\label{sc:Datasets}Data set}

We used observations of the Cosmic Evolution Survey field \citep[COSMOS;][]{Scoville_2007} obtained during the \Euclid\ performance-verification campaign (PV-005) in November 2023, covering $0.5\,\mathrm{deg}^{2}$. The data are part of Euclid Data Release 1 \citep[DR1;][]{DR1cite,DR1-TP001}. The observations include deep VIS imaging (\IE), NISP photometry (\YE, \JE, \HE), and NISP slitless spectroscopy, reaching depths comparable to those expected for the EDS. The analysis presented in this paper is based on the 116 blue-grism exposures, corresponding to a total integration time of $116\times549\,\mathrm{s}=63\,684\,\mathrm{s}$ ($17.7$\,h), all obtained with the same dispersion orientation because of the constant telescope roll angle. The PV-005 data set also contains 68 red-grism exposures distributed over four dispersion orientations, which will be the subject of future analyses.

\begin{figure*}
  \centering
  \includegraphics[width=\textwidth]{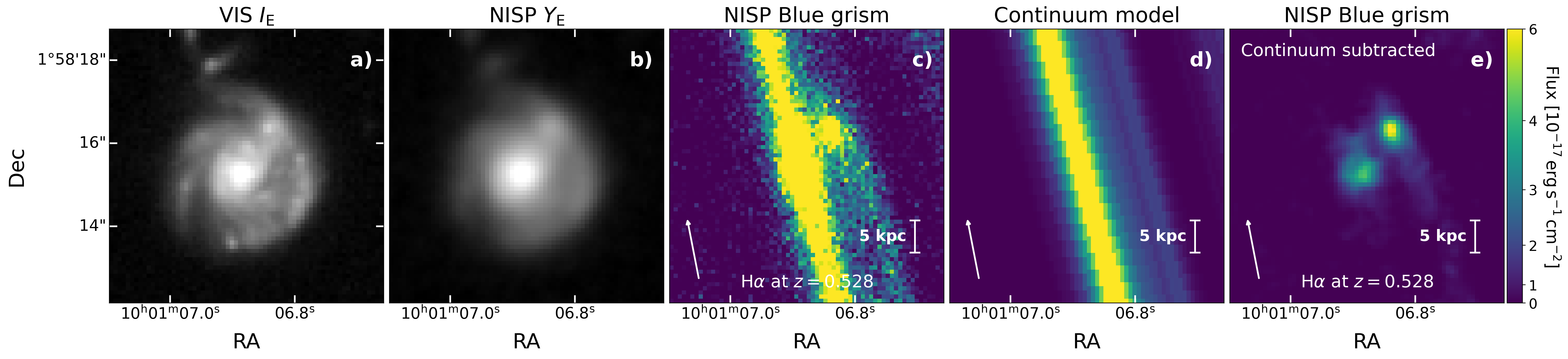}
\caption{Construction of an \Ha emission-line map for a face-on spiral galaxy at $z=0.528$ 
using 116 co-added blue-grism exposures. (a) \IE image. (b) \YE image. (c) \Ha map before continuum subtraction. (d) Model of the stellar continuum. (e) Continuum-subtracted and smoothed \Ha map. The arrows indicate the grism dispersion direction, and the vertical bars correspond to a physical scale of $5\,\mathrm{kpc}$.}
  \label{fig:median_filter}
\end{figure*}

We used three complementary catalogues, depending on the purpose of the analysis. The \Euclid\ spectroscopic extraction catalogue \citep[hereafter SPE;][]{Q1-TP007} provides the spectroscopic redshifts and 1D emission-line measurements used for the statistical analyses presented in Sect.~\ref{sect:6}. The corresponding 1D spectra, produced by the spectroscopic instrument reduction pipeline \citep[hereafter SIR;][]{DR1-TP023, DR1-TP024, Q1-TP006}, are also available for these sources and are shown for selected examples in Sect.~\ref{sc:exampleselmap}. Using SPE for both the redshifts and the reference line measurements enables a direct comparison between the 1D and 2D approaches within the same \Euclid\ data set. 

For the statistical analyses, we first selected the primary SPE redshift solutions with a redshift probability greater than 0.999, yielding \num{32235} sources from the full SPE catalogue of \num{87355} sources. We adopted the primary (highest-ranked) solution among the five redshift solutions provided by SPE. We then applied the following criteria:
\begin{itemize}
   \item $0.4<z_{\rm spec}<0.9$, corresponding to the redshift interval over which \Ha falls within the blue-grism wavelength coverage;
   \item ${\rm S/N}(\Ha)>3.5$;
   \item ${\rm S/N}>2.5$ for either the [\ion{S}{ii}]$\lambda\lambda6718,6733$ doublet or [\ion{O}{iii}]$\lambda5008$, the latter entering the blue-grism wavelength range at $z\gtrsim0.84$.
\end{itemize}
Here, the signal-to-noise ratios (S/N) are those reported in the SPE catalogue. The redshift selection retained 7644 sources, of which 6060 satisfied the \Ha criterion. Among these, 1106 satisfied neither additional line criterion, leaving 4954 sources. We cross-matched this SPE sample with the COSMOS2020 Classic catalogue \citep{Weaver_2022} using a maximum angular separation of \ang{;;0.36}. A COSMOS2020 counterpart was found for 4533 sources, while 421 remained unmatched. For the matched sources, we adopted the COSMOS2020 coordinates as the reference positions throughout this work, including for the extraction and reconstruction of the emission-line maps. The cross-match also provided the stellar masses and other photometric properties used in our analysis, notably in deriving the mass--metallicity relation presented in Sect.~\ref{sect:mzr}.

For the illustrative galaxies presented in Sect.~\ref{sc:exampleselmap}, we used the \citet{Khostovan2026} redshift catalogue, compiled from two decades of ground-based spectroscopy in the COSMOS field. These redshifts are particularly well suited to the redshift range probed by the blue grism and provide the accurate wavelength registration required to reconstruct individual emission-line maps. We selected star-forming galaxies with $0.4< z_{\rm spec}<0.9$ and $I_{\rm AB}\leq22.5$, using the rest-frame NUV$rJ$ colour--colour classification from COSMOS2020 \citep{Weaver_2022}. The remaining photometric quantities were also taken from COSMOS2020. SPE redshifts were used for some illustrative examples where explicitly stated.

\section{\label{sc:Gelsa}Methodology}

The capability to produce emission-line maps from \Euclid NISP data is implemented in the \texttt{GELSA} software\footnote{We used version 3.1 of \texttt{GELSA} in this work.} \citep[][Euclid Collaboration: Granett et al. in prep.]{granett_2026_23417008}, developed as part of the ELSA project.\footnote{ELSA: Euclid Legacy Science Advanced analysis tools, P.I. M. Talia, \url{https://elsa-euclid.github.io/}.} We started from the intermediate \texttt{DpdSirScienceFrame} data products generated by SIR within the \Euclid Science Ground Segment \citep[SGS;][]{DR1-TP033}. These products are background-subtracted NISP grism exposures that have undergone non-linearity correction and bad-pixel masking. We additionally applied a persistence mask, using a threshold of 30\,ADU to identify persistence-affected pixels. We refer to \citep[][]{DR1-TP003} for a description of persistence in NISP. Source localisation and flux measurements used the SIR astrometric, dispersion, and flux-calibration solutions, together with the data-quality information associated with each exposure.

The emission-line maps were constructed by resampling the NISP grism exposures onto a spatial RA--Dec grid at a fixed observed wavelength. The reconstruction does not require prior detection of emission-line sources and can be applied over larger areas, although we present cutouts centred on selected galaxies. The output grid was matched to the cutout extracted from the corresponding \Euclid photometric image, obtained from the MER mosaic \citep{DR1-TP009}. The emission-line and photometric images therefore share the same astrometric reference frame and pixel grid. The pixel scale is $\ang{;;0.1}$, compared with the NISP detector sampling of $\ang{;;0.3}$. The individual spectroscopic exposures were resampled and combined using the Drizzle algorithm \citep{Fruchter_2002}, adopting a pixel-shrink factor of 0.5, which sets the linear size of each drop to half of the native pixel, and 100 sub-pixel drops per detector pixel to sample its footprint. The finer output grid improves the spatial sampling of the dithered observations, recovering information from their sub-pixel offsets, but it does not improve the intrinsic angular resolution, which is determined by the NISP point-spread function (PSF, Sect.~\ref{sect:PSF}). The detector data-quality (DQ) information was propagated through the map construction. Pixels with non-zero DQ flags, non-finite values, or non-positive variances were excluded during resampling. The zero-order and persistence masks described above were also applied, such that masked pixels did not contribute to either the signal or the associated coverage map. The detector variance maps were corrected by the squared relative- and absolute-flux calibration factors and resampled using the same Drizzle footprints as the science data. Assuming independent exposures, their variances were summed and divided by the square of the corresponding normalisation when the maps were combined.

Stellar-continuum subtraction is a crucial step in constructing emission-line maps from grism exposures. In this work, we adopted a fast method that is applicable when the dispersion direction is the same in all exposures, which is the case for the COSMOS blue grism data. The continuum estimation was performed after resampling the exposures and co-adding them into the emission-line map. First, we rotated the emission-line map to align the dispersion direction with pixel rows. For each pixel, we estimated the continuum in a $400\,\AA$ interval (97 resampled pixels) using a sigma-clipped median. This resulted in a continuum image. We then rotated the continuum image back to the original frame and subtracted it from the co-added map. The presence of bright emission lines can bias the median estimate. To mitigate this effect, we used the spectroscopic redshift to predict the positions of \Ha, [\ion{N}{ii}]$\lambda6549$, and [\ion{N}{ii}]$\lambda6584$, and masked approximately $41\,\AA$ (10 pixels) on either side of the expected position of each line. The procedure is illustrated in Fig.~\ref{fig:median_filter}. In general, combining exposures with different dispersion orientations results in the superposition of spectral traces at different angles so that the continuum cannot be subtracted in this way.
This restriction is methodological and can be overcome with alternative continuum-subtraction procedures. Ongoing work is extending the analysis to NISP red-grism observations and to emission-line maps constructed from multiple orientations in the Euclid Deep Fields.
\begin{figure*}
    \centering
    \includegraphics[width=\textwidth]{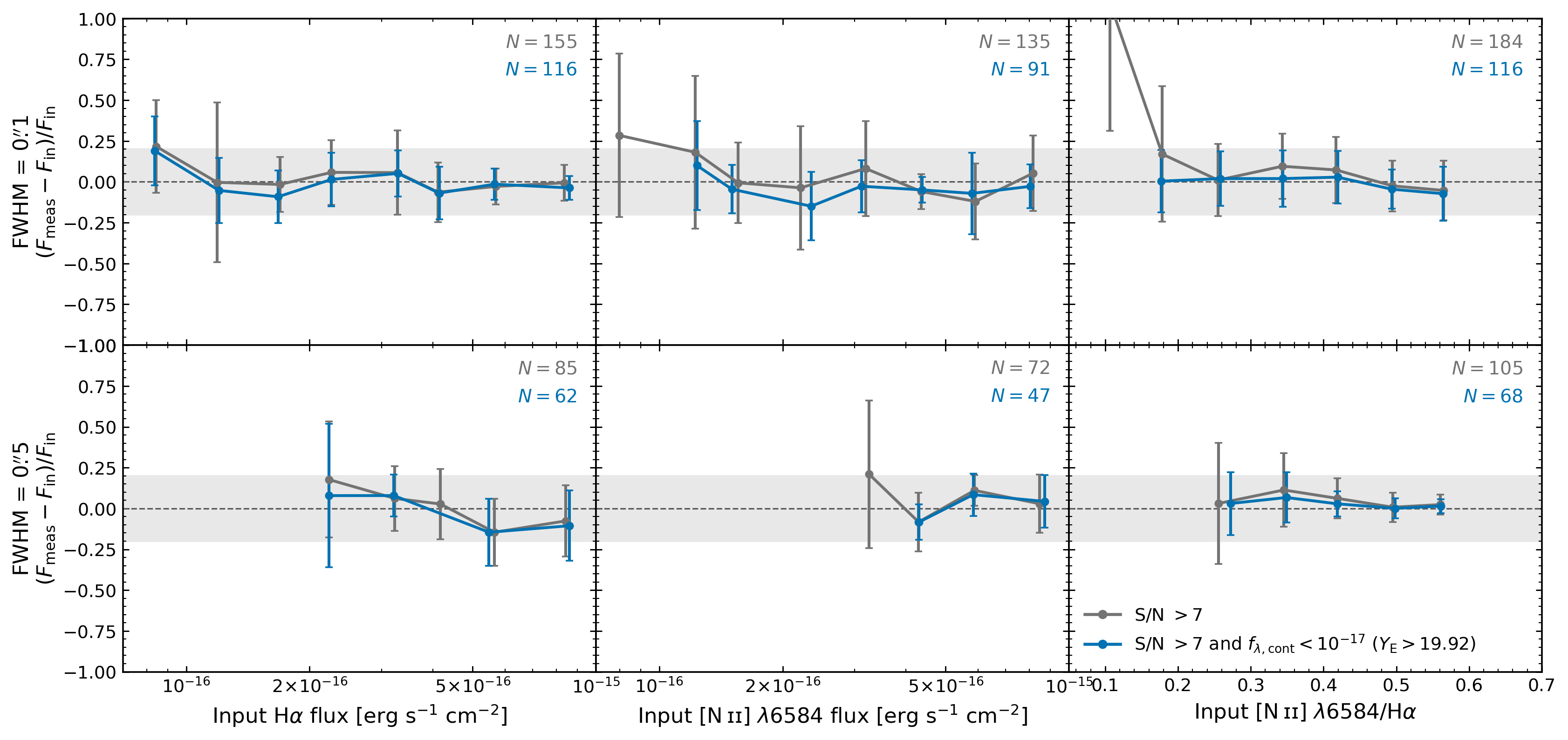}
    \caption{\emph{Top}: Flux recovery for sources with an intrinsic FWHM of \ang{;;0.1}. \emph{Bottom}: Same for sources with an intrinsic FWHM of \ang{;;0.5}. Curves show the median fractional residual for \Ha, [\ion{N}{ii}]$\lambda6584$, and their ratio [\ion{N}{ii}]$\lambda6584$/\Ha, with error bars showing the scaled MAD. Grey curves include sources with S/N$\,>7$, while blue curves additionally require $f_{\lambda,\mathrm{cont}}<10^{-17}\,{\rm erg\,s^{-1}\,cm^{-2}\,\AA^{-1}}$. For the line ratio, both lines are required to have S/N$\,>7$. The shaded region marks $\pm20\%$ recovery.}
    \label{fig:gelsa_in_out}
\end{figure*}

\begin{figure}
    \centering
    \includegraphics[width=0.5\textwidth]{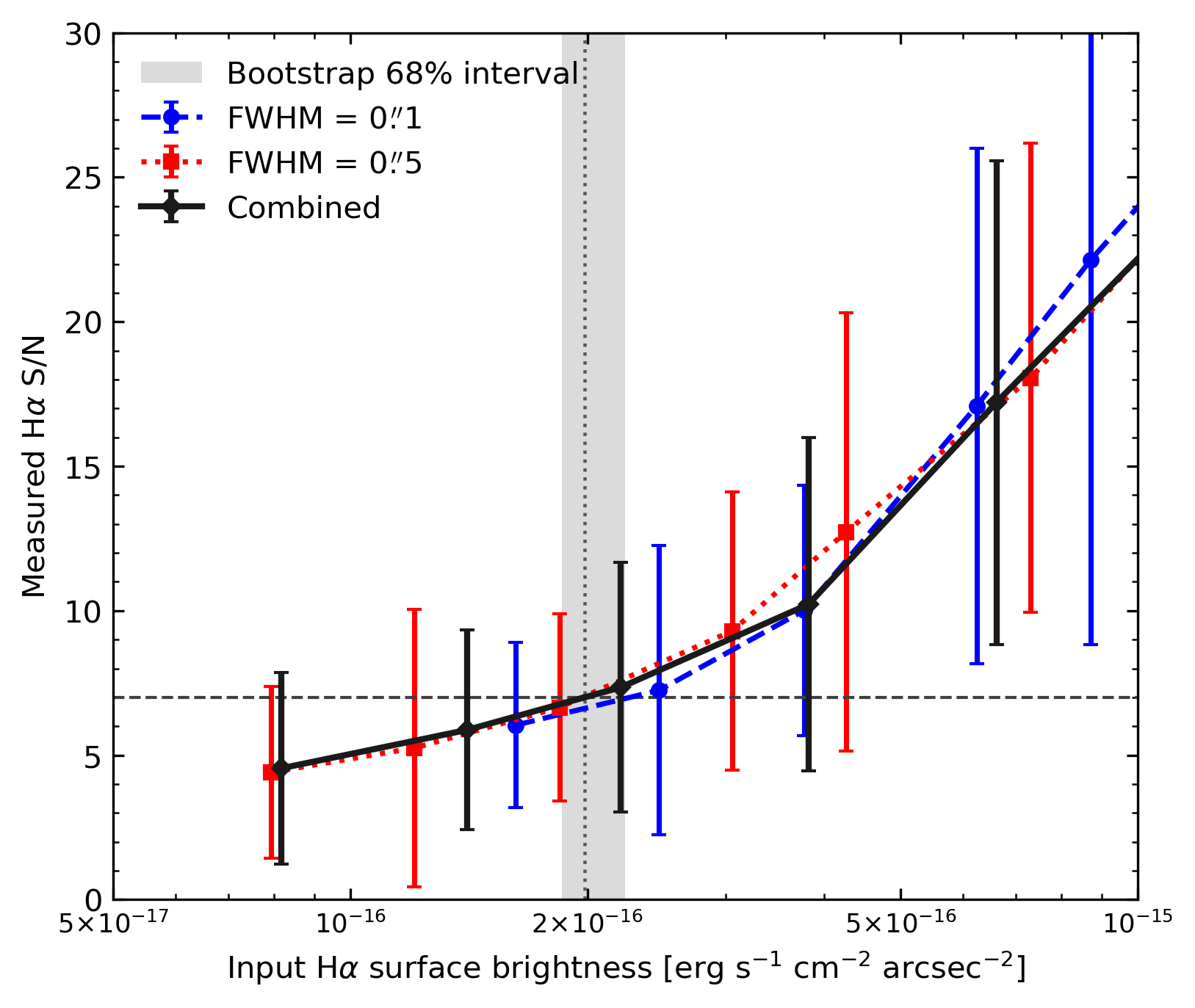}
\caption{Median \Ha S/N as a function of the input mean surface brightness within the PSF-convolved $R_{50}$. Blue dashed and red dotted curves show intrinsic FWHMs of \ang{;;0.1} and \ang{;;0.5}, respectively, while the black curve combines both sizes. Error bars show the scaled MAD, and the shaded region marks the 68\% bootstrap uncertainty on the ${\rm S/N}=7$ sensitivity.}
\label{fig:surface_brightness}
\end{figure}

\begin{figure}
    \centering
    \includegraphics[width=0.5\textwidth]{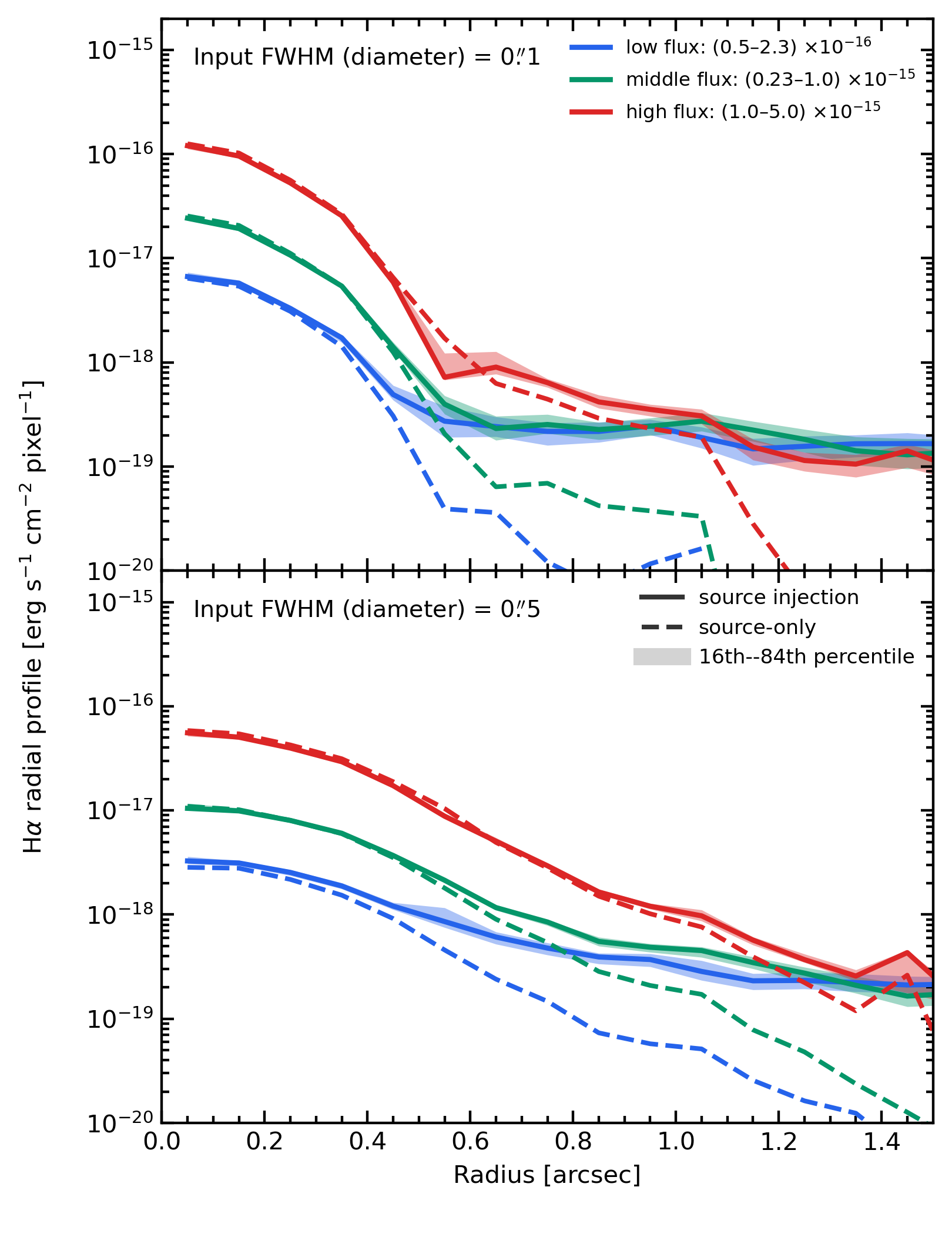}
    \caption{\emph{Top}: Recovered \Ha radial profiles for sources with an intrinsic FWHM of \ang{;;0.1}. \emph{Bottom}: Same for sources with an intrinsic FWHM of \ang{;;0.5}. Solid curves show the recovered profiles, while dashed curves show the source-only profiles processed using the same procedure. Colours indicate bins of injected \Ha flux, which are shown in the legend of the \emph{top} panel in ${\rm erg\,s^{-1}\,cm^{-2}\,\AA^{-1}}$, and shaded regions show the 16th--84th percentile range.
    }
    \label{fig:gelsa_in_out_radial_profile}
\end{figure}

At this stage, the resulting maps do not isolate emission at a single wavelength and should therefore not be interpreted as strictly monochromatic emission-line maps. Closely separated lines, such as \Ha and the [\ion{N}{ii}] doublet, appear as superposed spatial images displaced along the dispersion direction. This property can be used to detect and jointly model the \Ha--[\ion{N}{ii}] complex. Emission-line maps constructed from a broad line feature show the same effect: the spatial distribution of the emission is effectively convolved with the line profile. With a sufficient number of exposures obtained at different dispersion angles, it is possible to disentangle emission at individual wavelengths and construct monochromatic emission-line maps \citep{Griggio2026}. This development is ongoing (Euclid Collaboration: Rigamonti et al., in prep.). 

\section{Assessment of NISP capabilities for emission-line mapping}
\label{sc:assessment}

\subsection{Source-injection validation}
\label{sect:flux_calib}

We used source-injection simulations to test the recovery of emission-line fluxes and radial profiles. We added synthetic sources to real blue-grism exposures and processed them using the continuum-subtraction and measurement procedure adopted for the science analysis. The simulations included realistic detector noise, slitless contamination, persistence masking, cosmic rays, and instrumental residuals. We simulated four blue-grism exposures obtained at the same dispersion orientation, representative of one observation sequence. By comparison, the science analysis used 116 exposures, corresponding to 29 observations. The simulated redshifts were distributed over $0.4 < z< 0.9$, placing the \Ha--[\ion{N}{ii}] complex within the blue-grism wavelength range. The injected \Ha fluxes spanned $5\times10^{-17}$ to $5\times10^{-15}\,{\rm erg\,s^{-1}\,cm^{-2}}$, while the flat continuum ranged from $2\times10^{-19}$ to $1.5\times10^{-16}\,{\rm erg\,s^{-1}\,cm^{-2}\,\AA^{-1}}$, corresponding to $17<\YE<24.2$. The line-flux range was chosen to bracket the expected sensitivity for a single observation, while the continuum range sampled both bright and faint host galaxies to test the continuum-subtraction method. The injected [\ion{N}{ii}]$\lambda6584$/\Ha ratios were distributed between 0.05 and 0.60. The [\ion{N}{ii}]$\lambda6549$ flux was fixed to one third of [\ion{N}{ii}]$\lambda6584$, such that the latter represents three quarters of the total [\ion{N}{ii}] flux. We considered circular Gaussian sources with intrinsic FWHMs of \ang{;;0.1} and \ang{;;0.5}, spanning the range from effectively unresolved to significantly resolved sources at the NISP angular resolution and broadly covering the expected sizes of typical galaxies over the simulated redshift range. For each size, 576 source realisations were added at positions distributed across the 16 NISP detectors, giving 1152 synthetic sources in total. These simulations test the dependence on source extent, but not on morphology. Future studies will consider more realistic morphologies, including bulge--disc decompositions, elongated profiles, and different levels of clumpiness.

\begin{figure}
  \centering
\includegraphics[width=0.5\textwidth]{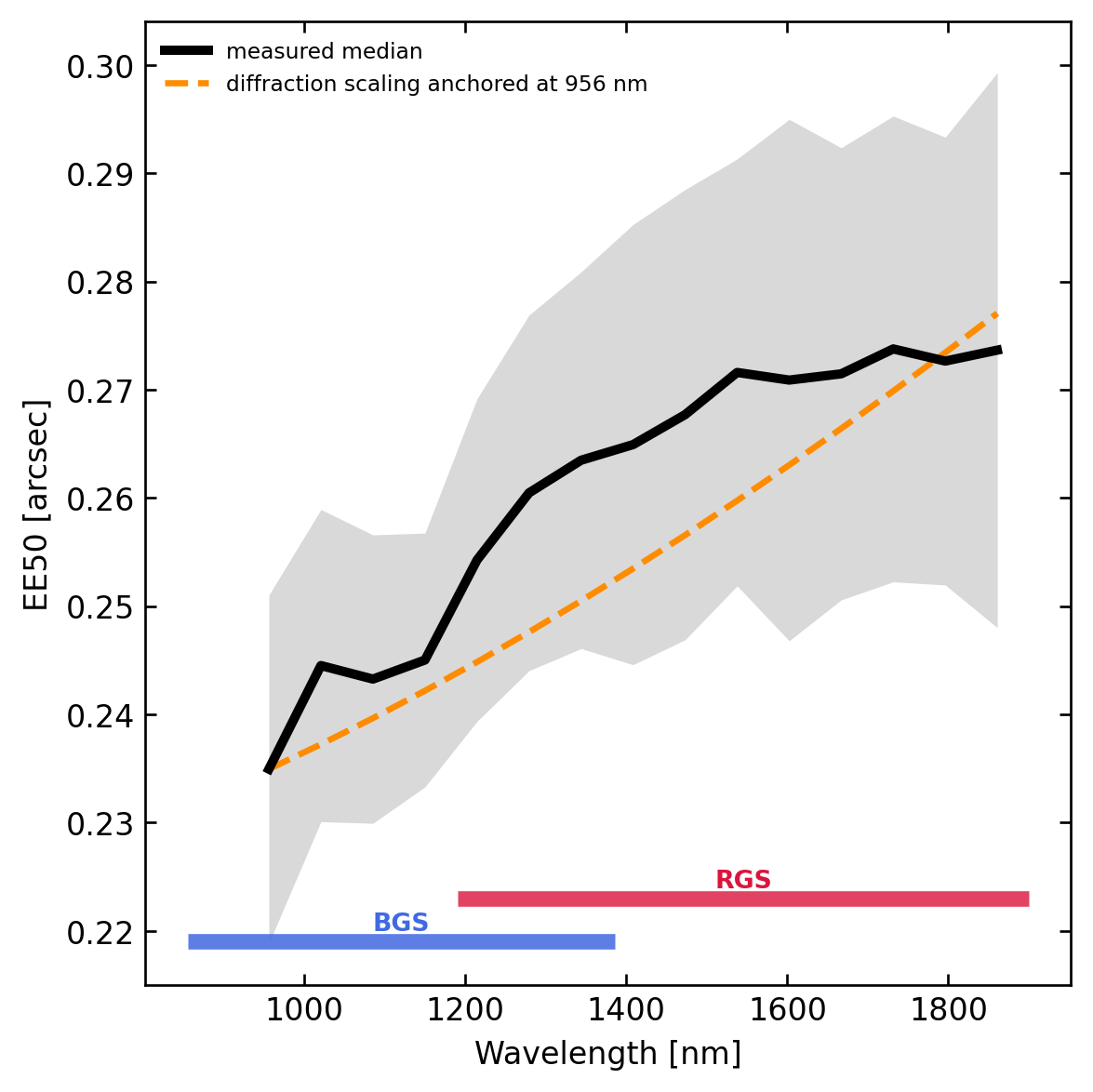}
\caption{Median EE50 versus wavelength for 295 MER point sources across the NISP focal plane, measured along the cross-dispersion direction in SIR science frames. The shaded region shows the $1\sigma$ dispersion among the 16 detectors. The orange dashed curve shows the expected diffraction scaling for \Euclid's 1.2-m aperture, anchored at the lowest wavelength; the horizontal bars mark the blue- and red-grism coverage.}
  \label{fig:med_sigma_psf}
\end{figure}

\begin{figure*}
  \centering
  \includegraphics[width=\textwidth]{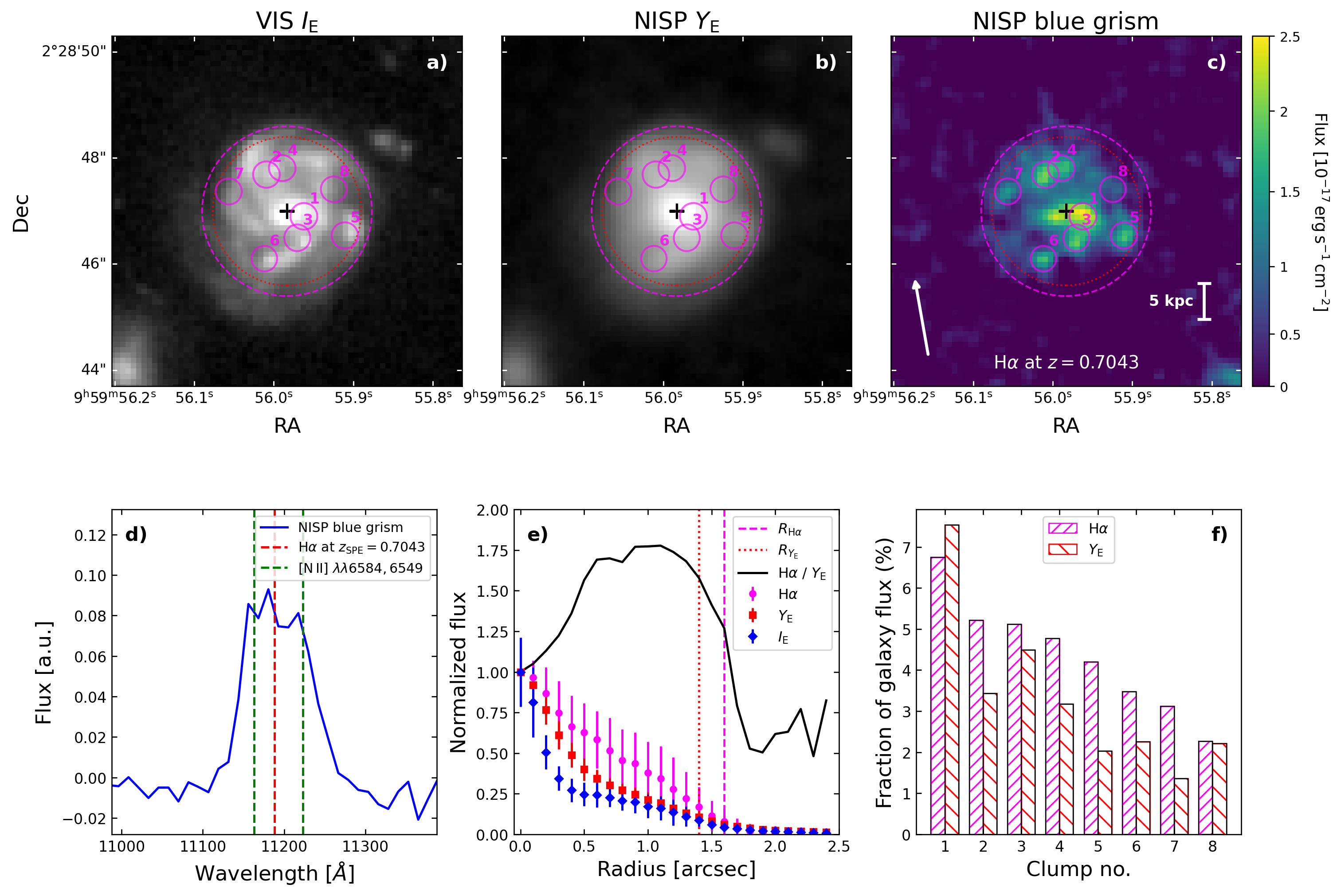}
\caption{\emph{Top}: (a) \IE image, (b) \YE\ image, and (c) \Ha emission-line map produced with the blue grism for a face-on galaxy at $z=0.7043$, using 116 stacked exposures. The black cross indicates the catalogue position. The dashed magenta and dotted red circles show the profile-based boundaries derived from the blue-grism and \YE\ radial profiles, respectively. The magenta apertures show the clumps identified with \texttt{DAOStarFinder} on the NISP blue grism \Ha map and fitted with 2D Gaussians. The arrows indicate the grism dispersion direction, and the vertical bars correspond to a physical scale of $5\,\mathrm{kpc}$. \emph{Bottom}: (d) continuum-subtracted 1D SIR spectrum around \Ha, where spatial features could be misinterpreted as the [\ion{N}{ii}] doublet. (e) Normalised radial profiles of the \IE, \YE, and \Ha emission. The black line shows the ratio between the normalised \Ha emission and \YE\ profiles. The vertical lines indicate the profile-based boundaries. (f) Fractional fluxes measured within the \Ha-defined clump apertures relative to the total blue-grism and \YE\ galaxy fluxes, shown by the unfilled, oppositely hatched magenta and red bars with black outlines, respectively. The apertures are ordered by decreasing integrated \Ha flux.}
  \label{fig:face-on}
\end{figure*}

The maps were continuum-subtracted using the method described in Sect.~\ref{sc:Gelsa}. We fitted the \Ha--[\ion{N}{ii}] complex with the three-component 2D Gaussian model used for the science measurements (Sect.~\ref{sect:mzr}). The relative line positions were fixed by their wavelength separations and the simulated redshift, while the three components shared the same fitted Gaussian geometry. The known injected size replaced the \YE morphology prior used for the real data, and the Gaussian widths were allowed to vary by the same multiplicative factors of 0.65--2.0 adopted in Sect. \ref{sect:mzr}. Fits with a major-axis width within $2\%$ of the upper bound were excluded because they generally corresponded to poorly constrained line decompositions. This criterion excluded $43\%$ of all injections: $39\%$ and $47\%$ for intrinsic FWHMs of \mbox{\ang{;;0.1}} and \ang{;;0.5}, respectively. Figure~\ref{fig:gelsa_in_out} shows the median fractional residual for \Ha, [\ion{N}{ii}]$\lambda6584$, and their ratio. For each line, we defined ${\rm S/N}=F/\sigma_F$, where $\sigma_F$ was derived from the covariance matrix of the three-component fit at the fitted source geometry. The pixel noise entering the covariance calculation was estimated from the median absolute deviation (MAD) of the model residuals. The grey curves include measurements with ${\rm S/N}>7$, the threshold adopted for consistency with the analysis of the real data in Sect.~\ref{sect:mzr}. The blue curves additionally require $f_{\lambda,\rm cont}<10^{-17}\,{\rm erg\,s^{-1}\,cm^{-2}\,\AA^{-1}}$, corresponding to $\YE>19.9$, to illustrate the effect of continuum brightness on the flux recovery. For the line ratio, both \Ha and [\ion{N}{ii}]$\lambda6584$ were required to satisfy the S/N criterion. The error bars show the robust scatter of the fractional residuals within each input bin, estimated as $1.4826$ times their MAD.

For the continuum-selected sample, the median residuals remain within approximately $5\%$. For the \ang{;;0.1} sources, the scaled MAD is $14\%$, $20\%$, and $17\%$ for \Ha, [\ion{N}{ii}]$\lambda6584$, and their ratio, respectively. The corresponding values for the \ang{;;0.5} sources are $24\%$, $21\%$, and $12\%$. Without the continuum cut, the scatter increases most clearly for [\ion{N}{ii}] and the line ratio, showing that stellar-continuum residuals remain an important limitation for bright sources with $\YE<19.9$. For fainter galaxies, the continuum is more effectively removed and has a substantially smaller effect on the recovered line fluxes.

We estimated the \Ha surface-brightness sensitivity as 

\begin{equation}
\left\langle\Sigma_{\Ha}\right\rangle_{R_{50}} = \frac{F_{\Ha}} {2\pi R_{50,\mathrm{obs}}^2}\, , 
\end{equation}
where $R_{50,\mathrm{obs}}$ is the PSF-convolved \Ha half-light radius measured from the source-only reference maps. As shown in Fig.~\ref{fig:surface_brightness}, the median ${\rm S/N}=7$ is reached at $(2.0\pm0.2)\times10^{-16}\, {\rm erg\,s^{-1}\,cm^{-2}\,arcsec^{-2}}$. The uncertainty here is the 16th--84th percentile bootstrap interval.

We also tested the recovery of the spatial distribution of the \Ha emission. Before stacking, we masked the expected positions of the two [\ion{N}{ii}] components along the dispersion direction, following the procedure adopted for the science analysis (Sect. \ref{sect:radial_profile}). We divided the sources into three equal-count bins of injected \Ha flux, median-stacked their 2D maps, and then measured the radial profiles from the stacks. Source-only maps were processed and stacked separately using the same procedure. These stacks provide an effective input reference that includes the broadening introduced by the instrument and map construction. Figure~\ref{fig:gelsa_in_out_radial_profile} compares the source-injection and source-only radial profiles. The profiles agree closely in their central regions, while the source-injection profiles reach a residual background floor at larger radii, particularly in the faintest flux bin. Relative to the source-only stacks, the recovered half-light radii agree within approximately $10\%$ in all flux and source-size bins. The simulations therefore show that the adopted procedure recovers the integrated line fluxes and central radial profiles, while measurements in the outer regions of faint sources remain limited by residual background structure.

\subsection{Wavelength and cross-dispersion mapping}
We evaluated the accuracy of the mapping between detector pixels and world coordinates (RA, Dec, $\lambda$) by comparing predicted and observed positions using multiple approaches: comparisons between \IE images and \Ha emission-line maps; analysis of stellar continua; and verification against SIR spectroscopic data. An offset in the cross-dispersion direction can remain after processing. This offset can be corrected through a calibration procedure that measures the trace position of stars, whose point-like nature enables accurate estimation of the offset between the expected trace, which is derived from the photometric images, and its observed location in the resampled 2D spectrum. SIR applies this procedure across all detectors, providing one correction value per detector. We observe a residual offset of approximately 0.3 of a pixel (\ang{;;0.1}), which could be further reduced by deriving corrections locally, using stars in the vicinity of the targets. In this work, we adopt the corrections provided by SIR \citep{DR1-TP024}.

\subsection{PSF determination}\label{sect:PSF}

The point-spread function (PSF) was quantified using the encircled-energy radius at 50\% (EE50) of point-like sources.
Gaussian profiles were fitted to the spatial traces of point-like sources identified in the \Euclid MER catalogue \citep{DR1-TP009},
and the EE50 value was derived from the fitted standard deviation using ${\rm EE50} = \sigma \sqrt{2 \ln 2} \, \times \, \ang{;;0.3}$. Figure~\ref{fig:med_sigma_psf} shows the median EE50 across the focal plane. The black curve represents the median per-detector value, while the shaded grey region indicates the standard deviation. The wavelength dependence follows the expected diffraction-limit scaling for \Euclid's 1.2m aperture. The median EE50 is \ang{;;0.265} and is consistent with the results presented in \citet{EP-Gillard}.
Interpreted physically, the median ${\rm EE50}=\ang{;;0.265}$ corresponds to a radius of $1.4$--$2.3\,{\rm kpc}$ over $0.4<z<1.8$. For a Gaussian profile, this is equivalent to an FWHM of \ang{;;0.530}, or $2.8$--$4.6\,{\rm kpc}$. This places the spatial resolution of the \Euclid\ emission-line maps between the typical $5$--$6\,{\rm kpc}$ FWHM achieved by ground-based integral-field unit surveys of star-forming galaxies, such as KROSS at $z=0.8$--$1.0$ \citep{Stott_2016,Harrison_2017} and KMOS3D at $z=0.6$--$2.7$ \citep{Wisnioski_2015,Wisnioski_2019}, and the roughly $1\,{\rm kpc}$ FWHM reached by HST/WFC3 grism observations at $z\approx1$ \citep{Nelson_2016} or adaptive-optics SINFONI studies at $z\approx2$ \citep{Forster_2009,Tacchella_2015}.

\begin{figure*}
  \centering
\includegraphics[width=\textwidth]{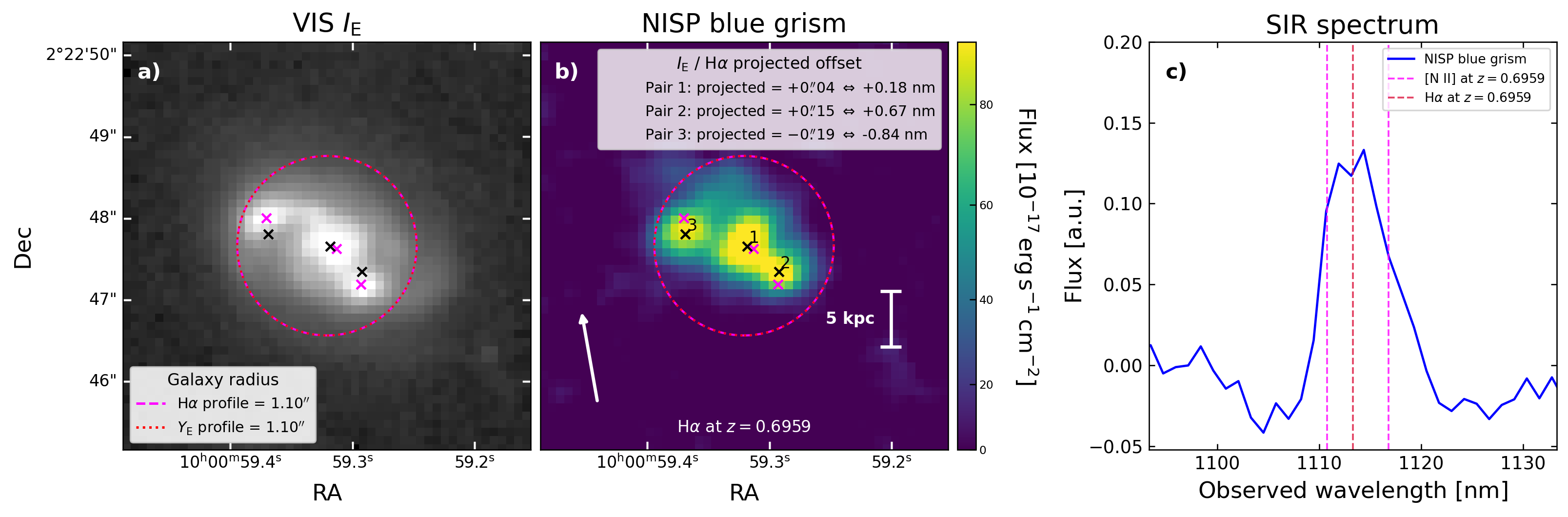}
\caption{Example of an inclined spiral galaxy illustrating how variations in line-of-sight velocity across the galaxy produce apparent spatial displacements in the reconstructed emission-line map ($z=0.6959$).
The \IE image (a) and continuum-subtracted blue-grism \Ha map (b) show three clumps identified with \texttt{DAOStarFinder}. Magenta and black crosses indicate the clump positions measured in the \IE image and emission-line map, respectively. The apparent offsets between the \IE\ and \Ha\ clump positions are consistent with Doppler-induced displacements along the grism dispersion direction. The arrow indicates the dispersion direction, the circles show the galaxy radii derived from the blue-grism and \YE profiles, and the white bar represents a physical scale of 5\,kpc. Panel (c) shows the continuum-subtracted 1D spectrum, with the expected positions of \Ha and the [\ion{N}{II}] doublet indicated by vertical dashed lines.}
  \label{fig:rotation_vel}
\end{figure*}

\section{Scientific capabilities of \Euclid emission-line maps}
\label{sc:exampleselmap}

Emission-line maps are a novel \Euclid data product that combines the spectral information from spectroscopy with the spatial information of direct imaging. Rather than focusing on a single scientific application, this section presents examples illustrating the range of analyses enabled by these maps. The aim is to demonstrate the different capabilities of the emission-line maps and to highlight the additional information that can be recovered compared with conventional 1D spectra. The galaxies presented in this paper are listed in Table \ref{tab:galaxies}.

\subsection{Spatially resolved analysis of a face-on galaxy} \label{onedblending} The principal advantage of emission-line maps is that they retain the spatial distribution of the line emission. In contrast, \Euclid 1D spectra are obtained by collapsing the 2D spectrograms along the cross-dispersion direction. Spatial information is lost in this process, although some signatures of the galaxy morphology remain imprinted in the resulting 1D spectrum. We first consider a face-on galaxy, for which the line-of-sight velocity gradients produced by rotation are limited and the emission-line map therefore provides a relatively direct view of the spatial distribution of the line emission. In Sect.~\ref{sc:rotation_vel}, we consider an inclined galaxy, where rotation introduces line-of-sight velocity shifts that are converted into apparent spatial displacements along the grism dispersion direction, distorting the morphology of the reconstructed emission-line map.

Figure~\ref{fig:face-on} shows an example of a resolved face-on galaxy at $z=0.7043$. Panels (a)--(c) show the \IE image, the \YE image, and the \Ha emission-line map obtained with the blue grism, respectively. Panel (d) shows the corresponding 1D spectrum extracted by SIR. Spatial structures within the galaxy can produce features in the collapsed spectrum that mimic emission lines. This effect can bias redshift determinations, line-flux measurements, and FWHM estimates, potentially leading to a misinterpretation of the physical properties of the source. In this example, residual spatial features occur close to the expected positions of the [\ion{N}{ii}]$\lambda\lambda6549,6584$ doublet, shown by the green lines in panel (d), and could therefore be incorrectly identified as [\ion{N}{ii}] emission associated with the galaxy.

Beyond breaking spatial and spectral degeneracies in the 1D spectrum, the spatial information recovered in the emission-line map enables us to study both the radial distribution of the emission and the properties of individual star-forming clumps. Radial profiles are particularly useful for investigating galaxy growth, bulge formation, and inside-out quenching \citep{Tacchella_2015,Tachella_2018,Matharu_2022,Estrada-Carpenter_2024}. Panel (e) presents the azimuthally averaged radial profiles of the \IE and \YE\ images and the \Ha emission-line map. The profiles are centred on the catalogue position and normalised to their central values. For the blue-grism and \YE\ profiles, the background was estimated from the approximately flat outer part of each profile and subtracted before determining the profile-based radius. We defined this radius as the first radial bin beyond which the background-subtracted profile remains below $10\%$ of its central value for three consecutive bins. The resulting radii are indicated by the vertical lines in panel (e) and by the dashed magenta and dotted red circles in panels (a)--(c). The black curve in panel (e) shows the ratio between the normalised blue-grism and \YE\ radial profiles. Its increase away from the centre indicates that the \Ha emission is less centrally concentrated than the \YE\ light.

We identified clumps in the \Ha map using \texttt{DAOStarFinder} of the \texttt{photutils} Python package. The search was restricted to the region within the radius measured from the blue-grism profile, and we adopted a fixed $3\,\sigma$ detection threshold relative to the locally measured background noise. Each detection was fitted with a 2D Gaussian, and the resulting apertures are shown in magenta in panels (a)--(c). The fitted clump positions were transferred to the \YE\ image using the astrometric solutions of the two images, allowing the same apertures to be used in both maps. Panel (f) shows the fractional fluxes measured within these apertures relative to the total blue-grism and \YE\ galaxy fluxes, with the magenta and red hatched bars, respectively. The clumps are ordered by decreasing integrated \Ha flux measured within their fitted apertures. For each image, the aperture flux was normalised by the corresponding total galaxy flux measured within the blue-grism radius. The \YE\ measurements include the underlying stellar-disc emission, which was not subtracted, and do not require a distinct clump detection in the \YE\ image. Possible spatial offsets between the \Ha and \YE\ emission are also not accounted for. A detailed assessment of the aperture-photometry method will be presented in future work. This example demonstrates how emission-line maps recover the spatial information lost in the collapsed 1D spectrum, enabling spatially resolved measurements of radial star-formation distributions and individual emission-line clumps.

\begin{figure*}
  \centering
  \includegraphics[width=\textwidth]{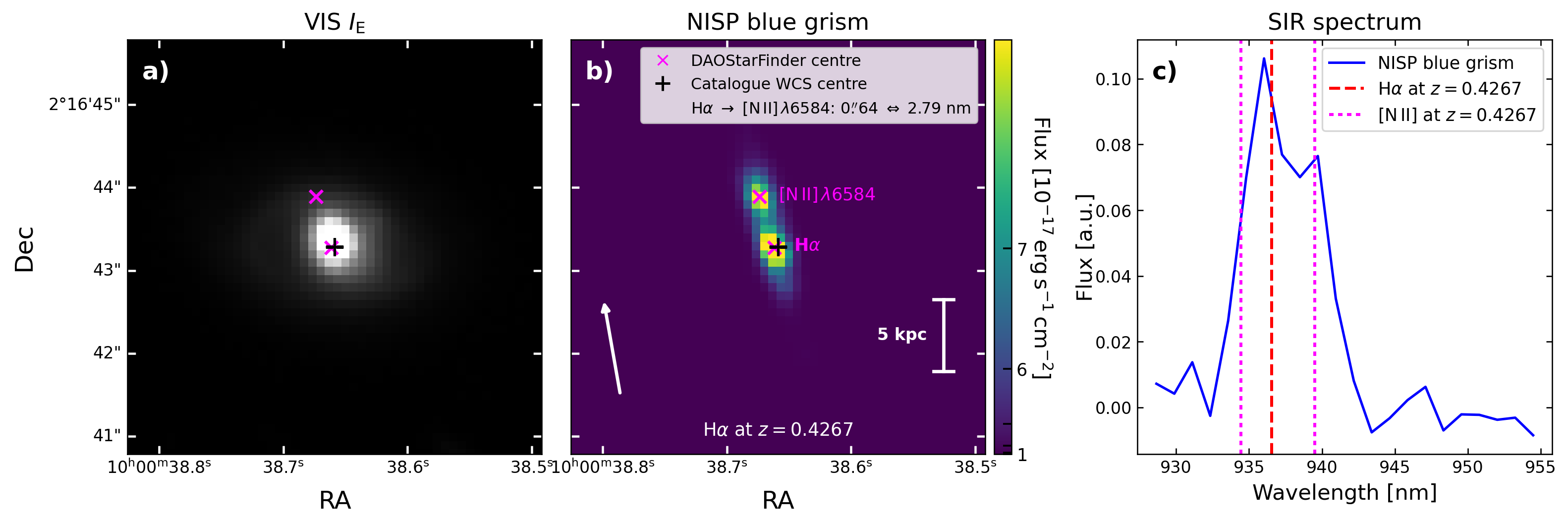} 
  \caption{Illustration of the spatial imprint of the \Ha--[\ion{N}{ii}] complex in a compact galaxy at $z=0.4267$. The \IE image (a) shows the stellar morphology, while the continuum-subtracted blue-grism map (b) shows two peaks aligned approximately with the dispersion direction. The two peaks, identified using \texttt{DAOStarFinder} on the median-filtered emission-line map, are associated with \Ha\ and [\ion{N}{ii}]$\lambda6584$. Magenta crosses indicate the \texttt{DAOStarFinder} centres, while the black cross marks the catalogue position. The measured separation between the \Ha and [\ion{N}{ii}]$\lambda6584$ peaks is \ang{;;0.64}, corresponding to approximately $2.79$\,nm along the dispersion direction. The arrows indicate the grism dispersion direction, and the vertical bars correspond to a physical scale of $5\,\mathrm{kpc}$. Panel (c) shows the continuum-subtracted 1D NISP spectrum, with the expected positions of \Ha and both [\ion{N}{ii}] lines indicated.}
  \label{fig:double-dot}
  \end{figure*}

\subsection{Kinematic signatures in emission-line maps}\label{sc:rotation_vel}

Emission-line maps resemble narrow-band images centred on a selected wavelength, but contain both spatially resolved spectral and kinematic information from the slitless spectroscopy. They therefore provide capabilities typically associated with integral-field spectroscopy. In particular, spatial variations in the line-of-sight velocity across a galaxy change the observed emission-line wavelength and therefore produce apparent spatial displacements along the grism dispersion direction. This effect was anticipated by \citet{Outini_2020} and has recently been exploited using JWST slitless observations of galaxies at $z=4$--6 \citep{Danhaive_2026}.

Figure~\ref{fig:rotation_vel} illustrates this effect for an inclined galaxy containing three bright clumps detected in both the \IE image and the blue-grism \Ha map. The offsets between the clump positions in the \IE image and \Ha map were projected along the grism dispersion direction. Clumps 2 and 3 show projected offsets of $+0\farcs15$ and $-0\farcs19$, corresponding to observed wavelength shifts of $+6.7$ and $-8.4$\,\AA, respectively. At the observed \Ha wavelength, their differential shift of $15.1$\,\AA\ corresponds to a projected velocity separation of approximately $407\,\mathrm{km\,s^{-1}}$. The opposite signs of the two shifts are consistent with the signature expected from a resolved velocity gradient across the galaxy.

If clumps 2 and 3 trace opposite sides of a rotating disc and are approximately symmetric about the systemic velocity, the measured separation can be interpreted as $v_{\rm rot}\sin i=204\,\mathrm{km\,s^{-1}}$. The MER catalogue gives a morphological position angle of $57^\circ$ and an axis ratio of $B/A=0.588$. Under the additional simplifying assumption of a geometrically thin circular disc, the axis ratio implies $i=54^\circ$, yielding an inclination-corrected rotational velocity of $v_{\rm rot}=252\,\mathrm{km\,s^{-1}}$. These values are intended to illustrate how kinematic information can be recovered from the spatial-spectral coupling of the slitless observations, rather than to provide a definitive measurement of the galaxy rotation velocity.

Several effects can contribute to the measured spatial offsets. The galaxy is clumpy and its intrinsic morphology may not be well described by a thin axisymmetric disc, while the \IE stellar continuum and \Ha emission do not necessarily trace identical regions, particularly in clumpy or dusty star-forming galaxies or systems containing an active galactic nucleus (AGN). Astrometric, wavelength-calibration, and redshift uncertainties can also contribute. Observations obtained at dispersion orientations separated by $180^\circ$ provide a way to distinguish these effects: a wavelength-induced displacement associated with a velocity gradient reverses direction, whereas an intrinsic spatial offset between the stellar continuum and line-emitting regions does not. Combining opposite orientations would therefore help disentangle morphology from kinematics and reduce sensitivity to common astrometric or wavelength zero-point errors. Such observations with different dispersion angles are available in the Euclid Deep Fields.

\begin{figure*}
  \centering
  \includegraphics[width=\textwidth]{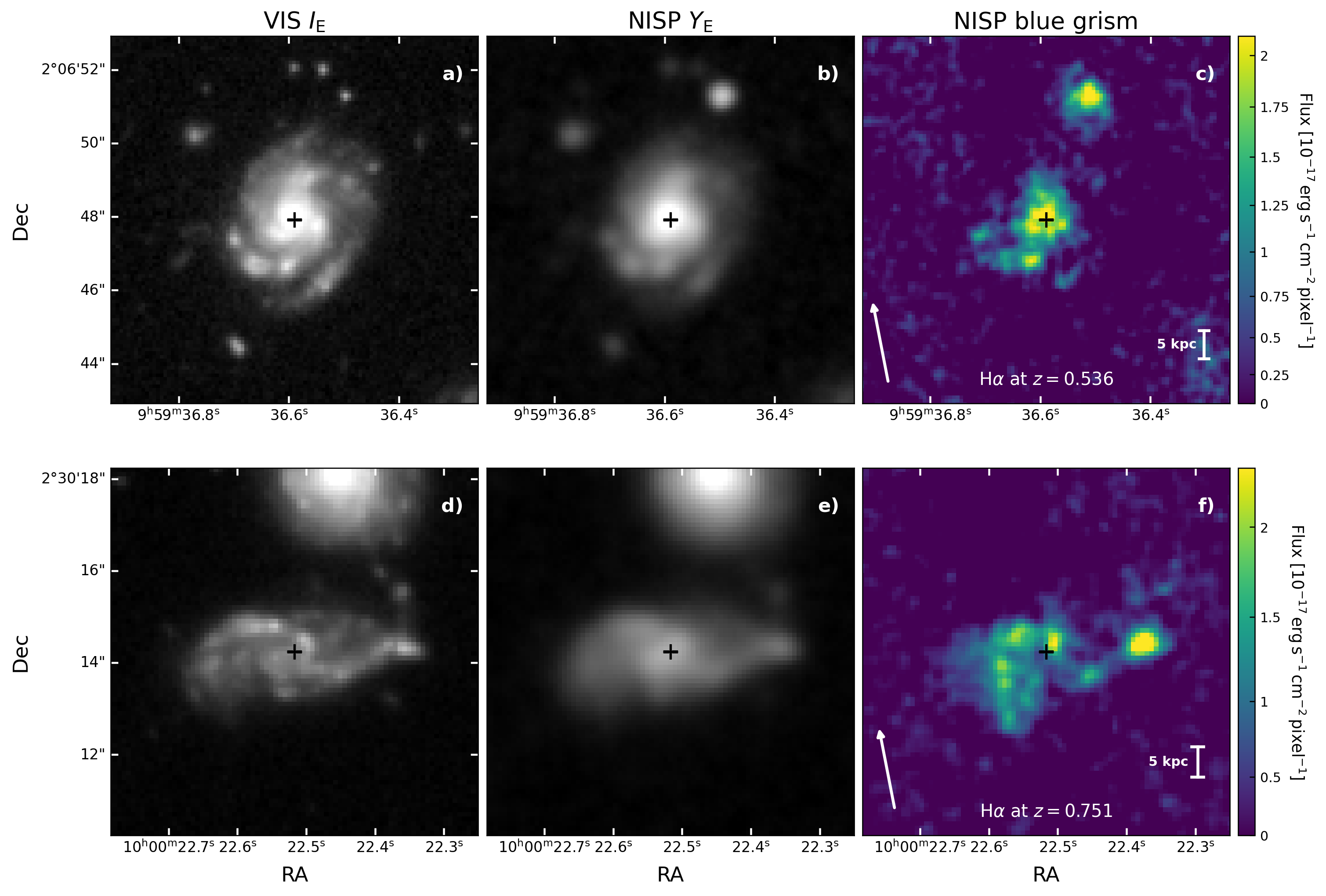}
\caption{Illustration of how emission-line maps reveal neighbouring or interacting galaxies by providing spectroscopic counterparts to sources identified in imaging. Panels (a) to (c) show the \IE image, the \YE image, and the \Ha emission-line map of a pair of galaxies at $z=0.5366$.
A compact source visible in the \IE and \YE\ images is even more prominent in the emission-line map, indicating that both galaxies are likely at the same redshift. Panels (d) to (f) show a system 
that appears to be interacting but whose components are actually at different redshifts: the upper source is not detected in the \Ha map constructed at $z=0.75$ and is instead associated with a foreground galaxy at $z=0.268$ included in \citet{Khostovan2026}. On the right of the image, the PSF-like feature exhibits enhanced \Ha emission, indicating an intense episode of star formation that could be associated with the galaxy. The arrows indicate the grism dispersion direction, and the vertical bars correspond to a physical scale of $5\,\mathrm{kpc}$.}
  \label{fig:Cluster_and_AGN}
\end{figure*}

\subsection{Deblending of the \Ha--[\ion{N}{ii}] complex}
\label{sect:deblending}

As discussed in Sect.~\ref{sc:Gelsa}, a wavelength (and redshift) must be specified to construct an emission-line map. When several emission lines are close in wavelength, their images are displaced to nearby positions along the dispersion direction and can overlap. This is particularly relevant for the [\ion{N}{ii}]$\lambda6549$--\Ha--[\ion{N}{ii}]$\lambda6584$ complex.  
Figure~\ref{fig:double-dot} illustrates this effect for a compact galaxy at $z=0.4267$. At this redshift, the rest-frame separations of $1.48$ and $2.07$\,nm between \Ha and [\ion{N}{ii}]$\lambda6549$ and [\ion{N}{ii}]$\lambda6584$, respectively, correspond to observed-frame separations of $2.10$ and $2.95$\,nm. Two peaks are independently detected in the emission-line map and are aligned approximately with the dispersion direction. They are identified as \Ha and [\ion{N}{ii}]$\lambda6584$ from their measured separation equivalent to $2.79$\,nm, consistent with the expected observed-frame separation at the resolution of the blue grism. 

With a single grism orientation, the image of a line complex cannot always be distinguished from genuine spatial clumps that happen to align with the dispersion direction. For each \Ha clump, however, the known wavelength separations of the triplet and the grism dispersion direction predict the positions at which its two [\ion{N}{ii}] counterparts should appear. A detected peak located at one of these predicted positions can therefore be flagged as a potential [\ion{N}{ii}] counterpart, whereas a peak that does not match the expected displacement is more likely to trace a distinct spatial clump. This identification remains ambiguous when the predicted position of an [\ion{N}{ii}] component coincides with an \Ha clump elsewhere in the galaxy. The \YE\ morphology prior and observations obtained at different grism orientations are therefore important for resolving such cases.

This example illustrates the information that emission-line maps can provide for deblending closely spaced emission lines. The quantitative deblending method adopted in this work, including the three-component model, the \YE\ morphology prior, and the comparison with SPE measurements, is presented in Sect.~\ref{sect:mzr}. The projected displacement depends on the grism orientation. Emission-line maps obtained at different orientations should therefore be modelled separately rather than combined without accounting for their different dispersion directions. Conversely, these complementary projections can provide additional constraints on the individual components and further improve the deblending, as will be explored in future work.

\begin{figure*}
  \centering
  \includegraphics[width=\textwidth]{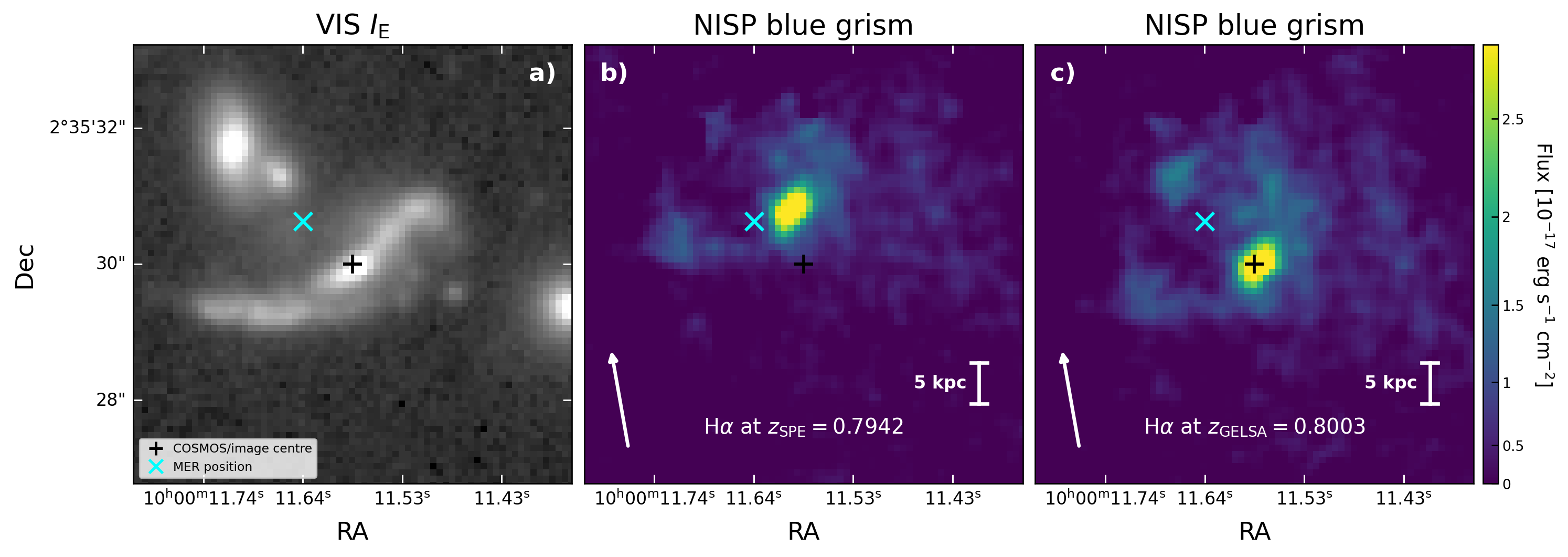}
\caption{Illustration of the effect of the adopted source position on the redshift measurement for an interacting system. Panel (a) shows the \IE image, with the COSMOS2020 position marked by the black cross and the MER position marked in cyan. Panel (b) shows the \Ha emission-line map reconstructed at $z_{\rm SPE}=0.7942$, measured from the spectrum extracted at the MER position. Panel (c) shows the map reconstructed at $z_{\rm GELSA}=0.8003$, estimated by matching the positions of identifiable emission-line structures to their counterparts in the direct images. The arrows indicate the grism dispersion direction, and the vertical bars correspond to a physical scale of $5\,\mathrm{kpc}$.}
  \label{fig:redshift_refinement}
\end{figure*}

\subsection{Spectroscopic characterisation of interacting systems}
\label{sc:systems_identification}

In 1D slitless spectra, spatial and spectral information are entangled, making extended, interacting, and crowded systems difficult to interpret. Emission-line maps preserve this spatial information and can therefore associate individual emission-line features with their photometric counterparts. Figure~\ref{fig:Cluster_and_AGN} presents two examples of this capability. Panels (a)--(c) show a compact source visible in the \IE and \YE images that is also detected in the \Ha emission-line map. The close spatial correspondence indicates that the source is likely at the same redshift as the target galaxy, since the emission-line map acts as a narrow-band image centred on \Ha.

Panels (d)--(f) show a second example in which the target galaxy is contaminated by the light of a foreground object at $z=0.2681$. Since the \Ha\ emission-line map is reconstructed at the target redshift ($z=0.751$), the foreground galaxy is not detected in the continuum-subtracted map. A compact PSF-like source is evident on the spiral arm on the right, providing an interesting interacting system for further study. These examples illustrate how emission-line maps can identify physically associated systems while rejecting unrelated foreground or background contaminants.

\subsection{Synergies with ground-based fibre spectroscopy}

A new generation of multi-object spectrographs, including WEAVE \citep{dalton_2016} and MOONS \citep{Cirasuolo_2020}, will obtain spectra for millions of pre-selected targets over the course of large surveys, with spectral resolving powers of $\lambda/\Delta\lambda \geq 2000$. In contrast, \Euclid slitless spectroscopy covers complete fields without spectroscopic target pre-selection, providing complementary observations of all sources within the overlapping fields. The finite size of fibres is a well-known source of aperture bias \citep[e.g.,][]{Brinchmann_2004,Kewley_2005,Gerssen_2012,Zahid_2013}. 
For galaxies comparable to or larger than the fibre diameter, a significant fraction of the emission-line flux can fall outside the aperture, complicating comparisons between fibre-based and spatially integrated measurements.

Emission-line maps constructed from grism spectroscopy can mitigate this problem by allowing fluxes to be extracted through apertures matched to the position and footprint of the ground-based fibres. This is particularly valuable when the facilities provide complementary diagnostics. For example, the WEAVE extragalactic surveys \citep{Shoko_2024} that overlap the \Euclid fields can provide H\,$\beta$, while \Euclid measures \Ha at $z>0.4$, enabling aperture-matched estimates of dust attenuation. This enables a synergy between \Euclid and the WEAVE/StePS survey \citep{Iovino_2023}. More generally, the same approach can provide consistent measurements of emission-line fluxes, star-formation rates, and gas-phase metallicities across facilities.

\subsection{Refining redshift measurements with emission-line maps}\label{sc:redshift_refinement} Although we used existing redshift measurements to build the emission-line maps presented in this paper, the spatial information preserved in slitless spectroscopic data can also be used to assess these measurements. In slitless spectroscopy, the wavelength and spatial position of a source are coupled along the dispersion direction. Consequently, either an error in the assumed wavelength of an emission line or an error in the adopted sky position of the galaxy can produce a displacement of the reconstructed emission-line map along this direction. Comparing identifiable structures in the emission-line map with their counterparts in the \Euclid VIS and NISP direct images can therefore reveal such inconsistencies and help assess the spectroscopic redshift.

Figure~\ref{fig:redshift_refinement} illustrates this effect for an extended interacting galaxy. The SPE pipeline provides two redshift solutions, $z=2.1590$ and $z=0.7942$. We adopted the secondary solution for consistency with the independent ground-based spectroscopic measurement, $z=0.8006$ \citep{Khostovan2026}. The \Ha emission-line map reconstructed at $z_{\rm SPE}=0.7942$, shown in panel (b), provides a plausible reconstruction of the line emission, but the emitting structures remain displaced along the dispersion direction from the corresponding structures in the \IE and \YE\ direct images.

In this case, the offset is related to the source position adopted for 1D SIR extraction and the subsequent SPE measurement. The MER position, shown in cyan in panel (a), is offset from the COSMOS2020 position adopted in this work, probably because the interacting system is treated as a single blended source in the MER catalogue. Since spatial position and wavelength are coupled along the dispersion direction, the component of this positional offset projected onto the dispersion axis propagates into the inferred wavelength and hence into the SPE redshift. This illustrates a limitation of 1D spectra for extended or blended systems: part of the spatial information is lost during the projection, and the resulting redshift can depend on the position adopted for the extraction.

We reconstructed the \Ha map for a series of trial wavelengths and compared the positions of identifiable emission-line structures with their counterparts in the \IE and \YE direct images. Panel (c) shows the reconstruction at $z_{\rm GELSA}=0.8003$, for which the line-emitting structures show the best spatial correspondence with the galaxy morphology. This value is consistent with the independent spectroscopic redshift of $z=0.8006$ reported by \citet{Khostovan2026}. Within the adopted \Ha identification, the procedure therefore corrects the apparent wavelength offset associated with the component of the positional error along the dispersion direction. This example shows how the spatial information retained in an emission-line map can reveal a bias caused by the adopted source position and help assess the redshift measurement.

This approach relies on identifying corresponding structures in the emission-line maps and direct images, and can therefore be limited by differences between the stellar-continuum and line-emitting morphologies. Observations at multiple dispersion orientations can help break the degeneracy between spatial position and wavelength, potentially allowing emission-line wavelengths to be constrained without relying on direct-image morphology. An automated approach exploiting this information is under development and has already been applied to trace arcs in strong-lensing systems \citep{DR1-Bergamini}.

\begin{figure}
  \centering
  \includegraphics[width=0.5\textwidth]{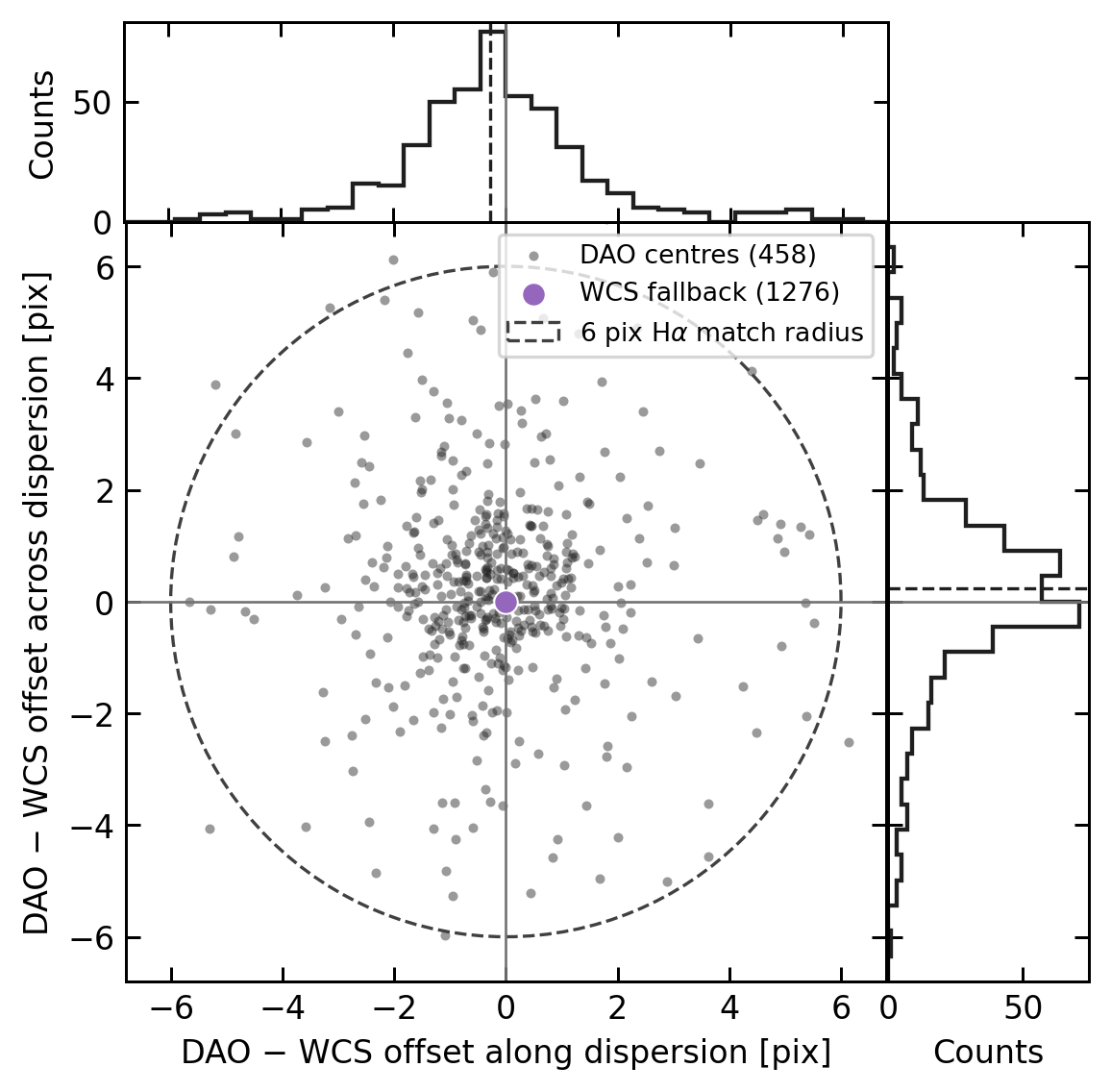}
\caption{Offsets between the refined \texttt{DAOStarFinder} and WCS source positions along and across the dispersion direction for galaxies at $0.4< z<0.9$. \texttt{DAOStarFinder} detections are initially associated with \Ha when located within six pixels of the WCS position, as indicated by the dashed circle. The selected position is subsequently refined to the local emission-line peak, so some final positions can lie slightly outside this radius. The marginal panels show the corresponding offset distributions.}\label{fig:WCSversusDAO}
\end{figure}

\section{Statistical analysis of emission-line maps}\label{sect:6}

\subsection{Mass--metallicity relation from emission-line maps}\label{sect:mzr}

The NISP spectral resolution implies that \Ha and [\ion{N}{ii}] are blended in slitless spectra, as shown in Sect.~\ref{sect:deblending}. The SPE pipeline performs a three-Gaussian fit to the 1D spectra to provide separated flux measurements for \Ha and [\ion{N}{ii}]. In this section, we test a complementary approach in which deblending is performed on the 2D emission-line maps by fitting the \Ha--[\ion{N}{ii}] complex with a three-component 2D Gaussian model.

\begin{figure*}
  \centering
  \includegraphics[width=\textwidth]{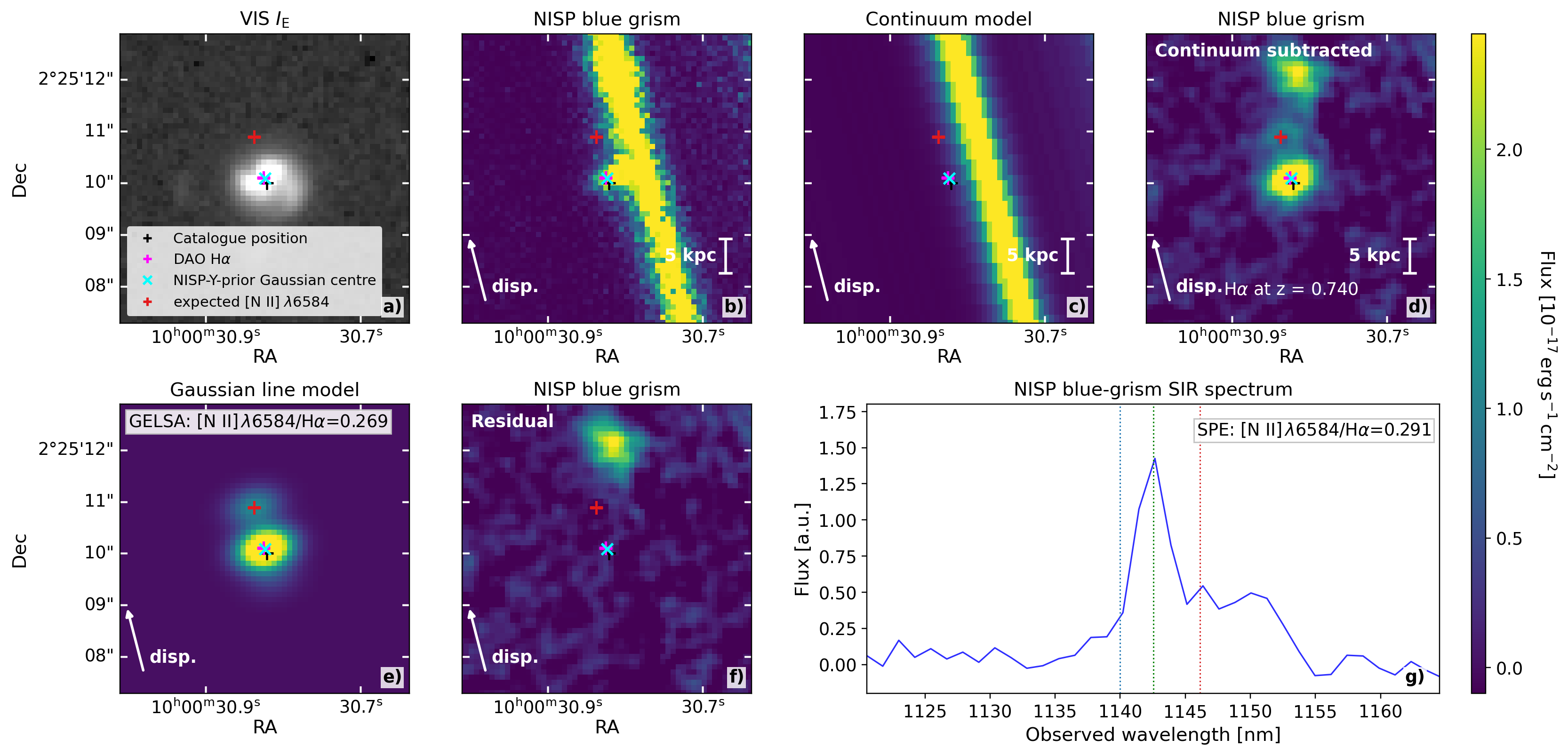}
\caption{Example of the extraction and deblending of the \Ha--[\ion{N}{ii}] complex for a galaxy at $z=0.7405$. (a) \IE image. (b) Blue-grism emission-line map before continuum subtraction. (c) Estimated continuum model. (d) Continuum-subtracted emission-line map, smoothed with a Gaussian kernel of $\sigma=1$ pixel. (e) Gaussian line model, with the spatial widths constrained by the \YE profile. (f) Corresponding residual. (g) Extracted SIR spectrum; the dotted lines indicate the expected positions of [\ion{N}{ii}] $\lambda6549$, \Ha, and [\ion{N}{ii}] $\lambda6584$. The [\ion{N}{ii}] $\lambda6584$/\Ha ratios measured by \texttt{GELSA} and SPE are reported in panels (e) and (g), respectively. The excess signal to the right of the complex is associated with a contaminating object visible at the top of the emission-line map and remains visible in the residual. The crosses mark the catalogue position (black), the \Ha position identified with \texttt{DAOStarFinder} on the continuum-subtracted map (purple), the centre of the \YE-profile-constrained fit (cyan), and the expected [\ion{N}{ii}] $\lambda6584$ position (red). The arrow indicates the dispersion direction, and the bar represents $5\,\mathrm{kpc}$.}
  \label{fig:residual}
\end{figure*}

\begin{figure}
  \centering
  \includegraphics[width=0.5\textwidth]{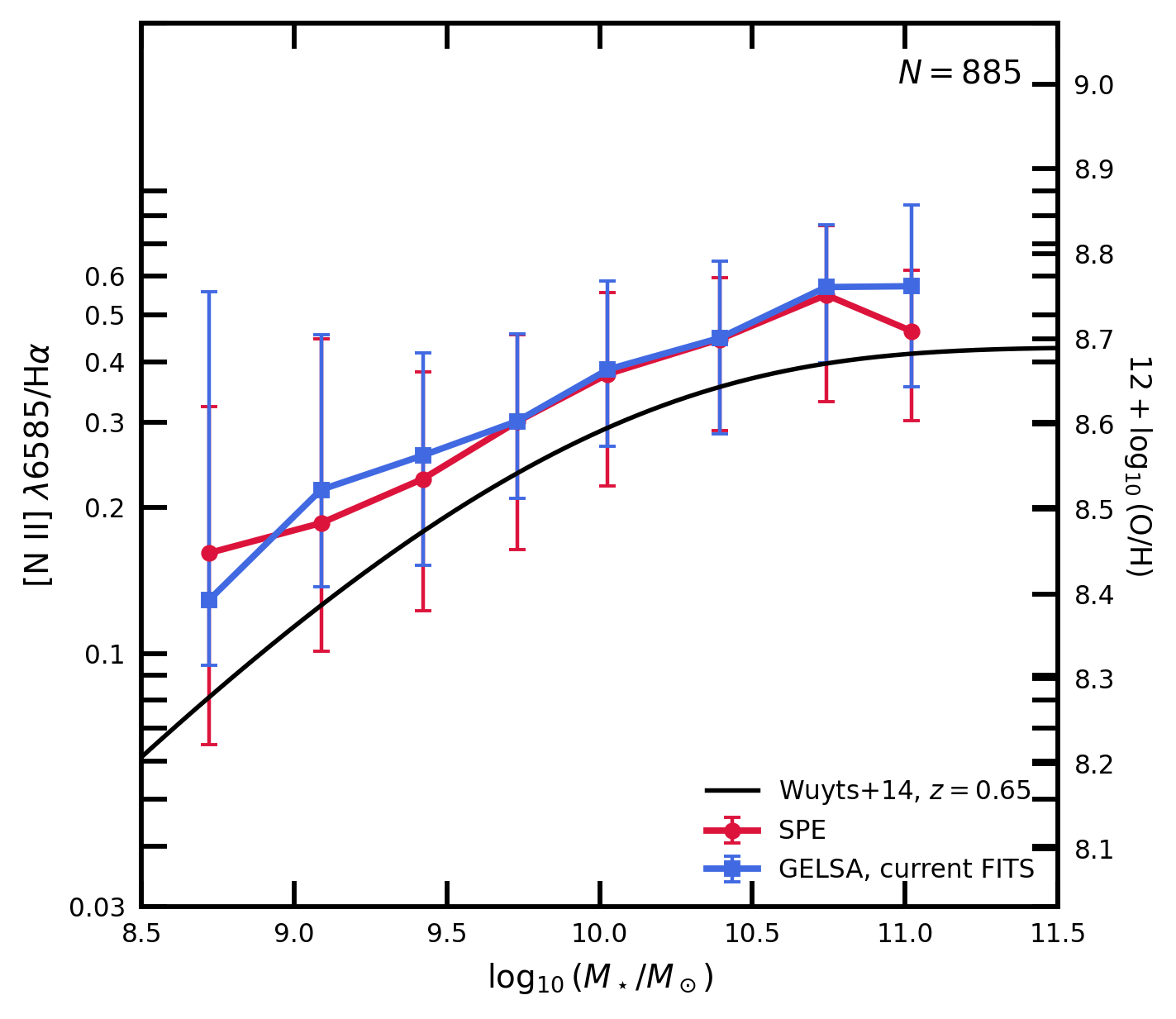}
\caption{Mass--metallicity relation derived from the SPE (red) and \texttt{GELSA} (blue) measurements for the same 885 galaxies at $0.4< z<0.9$. We required \Ha and [\ion{N}{ii}]$\lambda6584$ to be independently detected with \texttt{GELSA} ${\rm S/N}>7$. The curves show the median relations, while the error bars indicate the 16th--84th percentile ranges in each stellar-mass bin. The black curve shows the relation from \citet{Wuyts2014} at $z=0.65$.}
    \label{fig:mzr}
  \end{figure}

We selected objects from the Euclid DR1 observations and SPE catalogue in the COSMOS field following the criteria described in Sect.~\ref{sc:Datasets}. Among the 4533 sources with a COSMOS2020 counterpart, we further selected NUV$rJ$-classified star-forming galaxies with clean COSMOS2020 photometry and a galaxy classification (\texttt{FLAG\_COMBINED}=0 and \texttt{lp\_type}=0). These additional criteria removed 1546 sources, leaving 2987 galaxies at $0.4<z<0.9$. In this redshift range, the \Ha--[\ion{N}{ii}] complex is observed with the blue grism. As in the previous analyses, all blue-grism exposures have the same dispersion orientation. This is important for the present implementation because it allows us to estimate and subtract the continuum consistently along the dispersion direction directly from the emission-line maps, as introduced in Sect.~\ref{sc:Gelsa}. Since the blue-grism exposures share the same dispersion orientation, the expected offsets between \Ha and the two [\ion{N}{ii}] components project along a fixed direction in the emission-line maps, as illustrated in Fig.~\ref{fig:double-dot}. The relative positions of the triplet can therefore be predicted from the redshift and fitted with a single model. For the red-grism exposures, which have several orientations, the deblending must be performed separately for each orientation.

For each source, we first constructed the emission-line map by resampling and co-adding the 116 individual blue-grism exposures on a common RA--Dec grid. The continuum was then estimated directly from the co-added map. When performing the continuum estimation, we masked the emission lines in a region extending $\ang{;;1}$, or approximately $41\,\AA$, on either side of the expected positions of \Ha and the two [\ion{N}{ii}] components, as predicted from the primary SPE redshift solution. The continuum was estimated from the remaining pixels using the procedure described in Sect.~\ref{sc:Gelsa}. The resulting continuum model was subtracted from the original map without clipping the residuals, thereby preserving both positive and negative fluctuations. 

\begin{figure*}
  \centering
  \includegraphics[width=\textwidth]{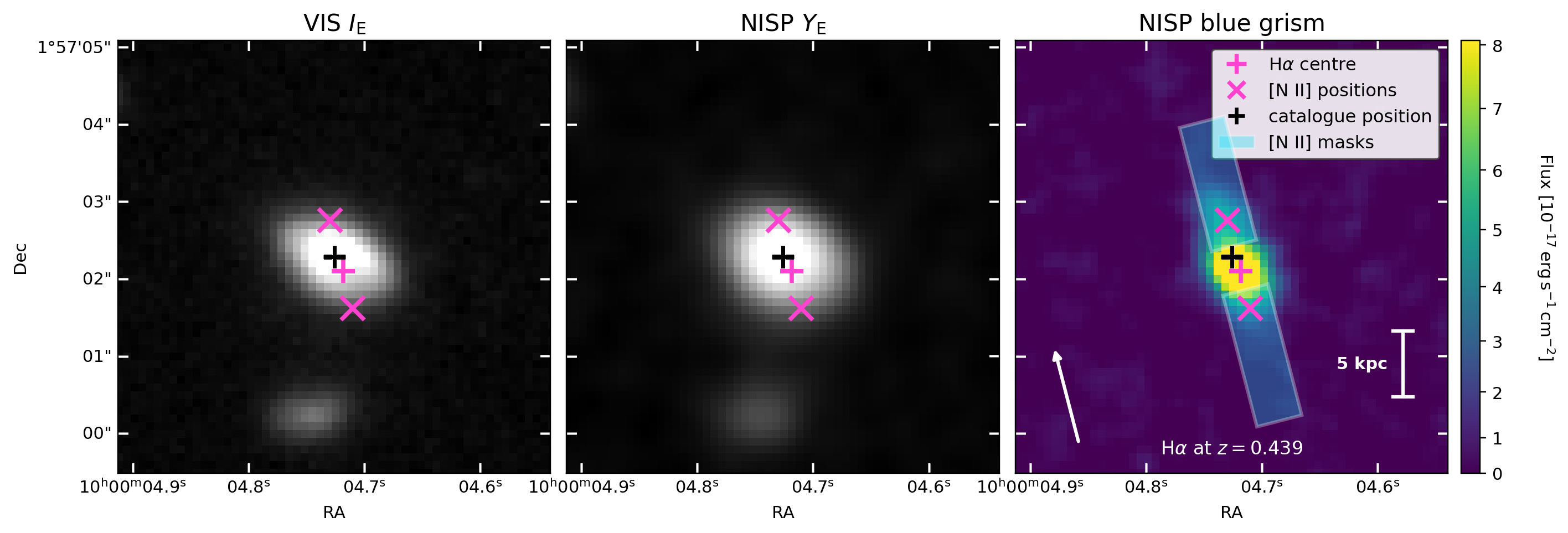}
\caption{Example of the \IE image, \YE image, and \Ha emission-line maps for a galaxy at $z=0.4390$. The black cross indicates the catalogue position, while the magenta symbols show the \Ha centre and the expected [\ion{N}{ii}] positions. The blue regions indicate the masks applied when measuring the \Ha radial profile. The arrow shows the grism dispersion direction, and the vertical bar corresponds to a physical scale of 5~kpc at the adopted redshift.}
\label{fig:smallpacman}
\end{figure*}

We smoothed the continuum-subtracted emission-line map with a Gaussian kernel with width $\sigma=1$ pixel. To obtain an initial centre, we applied \texttt{DAOStarFinder} to this map with a detection threshold of $2.3\,\sigma$ and a detection kernel with an FWHM of one pixel. When several detections were found, we selected the strongest source within six pixels of the expected \Ha position. This maximum separation was adopted to ensure that the selected emission-line peak remained spatially consistent with the position of the source in the photometric data. When no detection satisfied this criterion, we retained the catalogue position. The offsets between the catalogue and \texttt{DAOStarFinder}-based positions are shown in Fig.~\ref{fig:WCSversusDAO}.

The \Ha--[\ion{N}{ii}] complex was then fitted using the primary \textit{Euclid} spectroscopic-redshift solution to predict the relative positions of \Ha, [\ion{N}{ii}]$\lambda6549$, and [\ion{N}{ii}]$\lambda6584$. The three components were fitted simultaneously using 2D Gaussian models, with their relative positions fixed by the expected separations along the dispersion direction. The centre was allowed to vary by $\pm0.5$ pixel along each axis from the initial position. The [\ion{N}{ii}]$\lambda6549$ amplitude was fixed to 1/3 of the [\ion{N}{ii}]$\lambda6584$ amplitude \citep{Acker1989}, as adopted by the SPE flux-measurement pipeline.

The Gaussian widths along and across the dispersion direction, $\sigma_u$ and $\sigma_v$, were constrained using the morphology measured from the corresponding \YE\ image, but were allowed to vary by multiplicative factors of 0.65--2.0. These bounds were adopted to prevent unrealistically narrow or broad solutions driven by noise, but may bias the measurements or selection of galaxies whose \Ha\ emission is substantially more compact than their \YE\ continuum. The impact of this assumption will be investigated in future work. A constant residual-background component was also included in the fit. The Gaussian templates were normalised to unit integral, so that their fitted amplitudes correspond directly to integrated line fluxes. We finally computed the residual map as a diagnostic of the continuum subtraction and deblending quality. We illustrate the procedure in Fig.~\ref{fig:residual}. 

\begin{figure*}
  \centering
  \includegraphics[width=\textwidth]{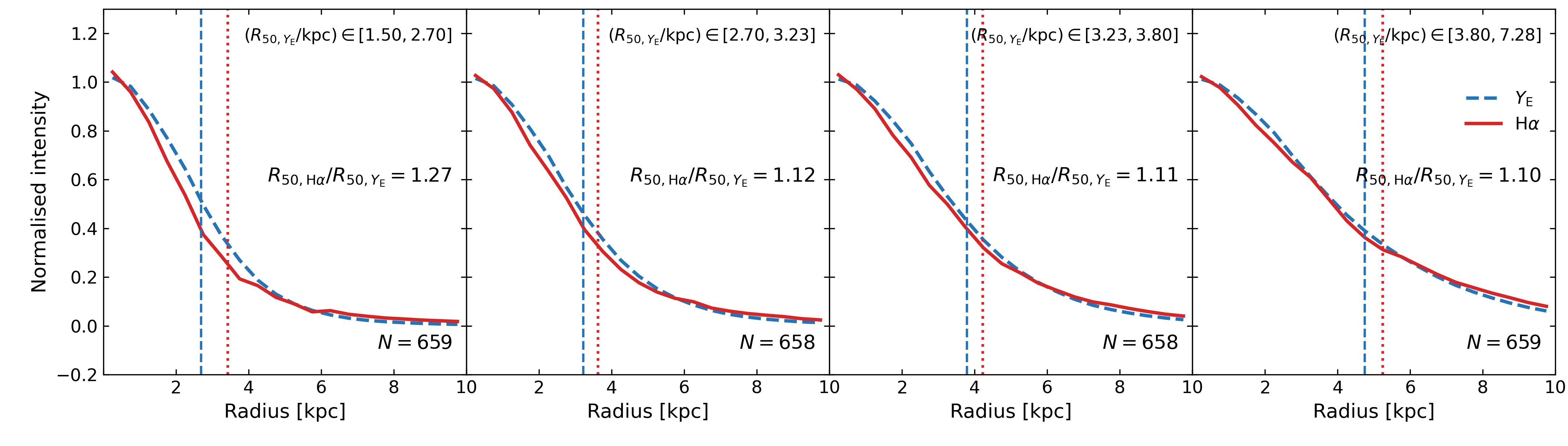}
  \caption{Normalised radial profiles measured from the median-combined NISP \YE\ (blue) and \Ha\ (red) maps for 2634 galaxies at $0.4<z<0.9$. The sample was divided into four equal-number bins according to the individual \YE\ $R_{50}$. Before stacking, the maps were normalised by the \YE\ flux, and the \YE\ images were PSF-matched to the \Ha\ resolution. The \Ha\ maps were centred using \texttt{DAOStarFinder}, with the predicted [\ion{N}{ii}] regions masked. Vertical lines indicate the corresponding $R_{50}$ measurements, with their ratios reported in each panel.}
\label{fig:radial_profile_stack}
  
\end{figure*}

We then used the \Ha and [\ion{N}{ii}]$\lambda6584$ fluxes to derive the mass--metallicity relation (MZR) from the [\ion{N}{ii}]$\lambda6584$/\Ha ratio, hereafter $N_2$. We converted this ratio into oxygen abundance using the $N_2$ calibration of \citet{Pettini2004},
\begin{equation}
12+\log_{10}({\rm O/H}) =
8.90+0.57\log_{10}(N_2)\,.
\label{eq:n2_metallicity_calibration}
\end{equation}
This calibration is also used for the metallicity scale shown on the right-hand axis of Fig.~\ref{fig:mzr}.

We computed the \texttt{GELSA} line S/N ratios following the procedure described in Sect.~\ref{sect:flux_calib}. We required both emission lines to be independently detected in the maps with
\begin{equation}
\frac{F_{\Ha}}{\sigma(F_{{\rm H}\,\alpha})}>7
\qquad {\rm and} \qquad
\frac{F_{[\ion{N}{ii}]\,\lambda6584}}
{\sigma(F_{[\ion{N}{ii}]\,\lambda6584})}>7\,.
\label{eq:line_snr_selection}
\end{equation}
The S/N criterion was therefore applied separately to \Ha and [\ion{N}{ii}]$\lambda6584$, rather than to their flux ratio. This is the same criterion adopted in the source-injection analysis and ensures that both measurements entering the metallicity indicator are individually detected with sufficient significance. The purpose of this comparison is to validate the 2D line-ratio recovery for a common selected sample, rather than to derive an unbiased mass--metallicity relation for the underlying galaxy population. 
After applying the line-detection criteria, we obtained a sample of 980 galaxies. Excluding fits for which either Gaussian width lay within $2\%$ of its upper bound removed 95 sources ($9.7\%$), leaving a final sample of 885 galaxies. Figure~\ref{fig:mzr} compares the MZR obtained from the \Euclid SPE Gaussian-fit measurements with that derived from the 2D emission-line-map fitting for the same galaxies. 
The two methods recover similar MZRs, indicating overall consistency between the 1D and 2D deblending approaches. They provide complementary measurements and an important cross-check: SPE measures the integrated line emission from the 1D spectrum, while \texttt{GELSA} preserves the spatial information and can distinguish emission associated with different regions or contaminating sources in the 2D maps. Both measurements show a small offset relative to the relation of \citet{Wuyts2014}. This difference may partly arise from the requirement of significant detections of both \Ha\ and [\ion{N}{ii}]$\lambda6584$, which preferentially selects galaxies with higher [\ion{N}{ii}]$\lambda6584$/\Ha\ ratios, particularly at low stellar masses. A quantitative assessment of this selection bias is deferred to future work.

\subsection{Radial profile analysis of \Ha emission-line maps}
\label{sect:radial_profile}

We finally used the emission-line maps to investigate the radial distribution of \Ha emission. 
For this analysis, we selected a sample of galaxies and stacked their emission-line maps to measure the average radial profiles. 
From the sample defined in Sect.~\ref{sc:Datasets}, 2634 galaxies at $0.4 < z < 0.9$ had usable maps and \YE flux normalisations.
For the stacking analysis, we did not impose the \Ha and [\ion{N}{ii}] S/N criteria that were adopted for the mass--metallicity relation.
We projected the \YE\ and \Ha\ maps onto a common physical grid sampled at $0.5\,\mathrm{kpc\,pixel^{-1}}$ using the redshift of each galaxy. Since \Ha\ falls within the \YE\ passband over most of the redshift range considered, we focus on the comparison between \Ha\ and \YE\ radial profiles, leaving the analysis of the \IE, \JE, and \HE\ profiles to future work. Before stacking, both maps of each galaxy were normalised by their integrated \YE\ flux, providing a common continuum normalisation. The \YE\ images were aligned on their catalogue positions. For \Ha, we used the source centre identified with \texttt{DAOStarFinder}, retaining the catalogue-based position when no reliable detection was obtained. To reduce contamination from the [\ion{N}{ii}] doublet, we masked two rectangular regions aligned with the dispersion direction and covering the expected [\ion{N}{ii}]$\lambda6549$ and [\ion{N}{ii}]$\lambda6584$ positions before stacking, as illustrated in Fig.~\ref{fig:smallpacman}.

We measured an individual \YE\ $R_{50}$ for each galaxy and divided the sample into four approximately equal-number bins as follows: $R_{50,\YE}\in[1.50,2.70]\,{\rm kpc}$, $R_{50,\YE}\in[2.70,3.23]\,{\rm kpc}$, $R_{50,\YE}\in[3.23,3.80]\,{\rm kpc}$, and $R_{50,\YE}\in[3.80,7.28]\,{\rm kpc}$. Before stacking, we matched the \YE\ PSF, with $R_{\rm EE50}=\ang{;;0.22}$ \citep{EuclidSkyNISP}, to the broader emission-line-map PSF of $R_{\rm EE50}=\ang{;;0.265}$ measured in Sect.~\ref{sect:PSF}. Within each bin, the PSF-matched \YE\ and \Ha\ maps were median-stacked pixel by pixel before measuring their azimuthally averaged radial profiles.

Figure~\ref{fig:radial_profile_stack} shows the resulting \YE\ and \Ha\ profiles, normalised to their central intensities for visualisation. The vertical lines indicate the $R_{50}$ measurements derived from the stacked profiles. The \Ha\ emission remains more extended than the PSF-matched \YE\ continuum in all four size bins. From the smallest to the largest \YE\ size bin, we measure $R_{50,\Ha}/R_{50,\YE}=1.27$, $1.12$, $1.11$, and $1.10$, respectively. The relative \Ha\ extension is strongest among the smallest galaxies, although the modest amplitude of this trend and the simplified Gaussian PSF matching mean that it should not yet be interpreted physically. Overall, our measurements agree with previous studies finding that \Ha emission is typically more extended than the stellar continuum \citep[e.g.,][]{Nelson_2016,Wilman_2020}. This analysis demonstrates that emission-line maps retain spatial information that can be used to investigate the distribution of recent star-formation activity. A detailed study of intrinsic, PSF-corrected sizes and their dependence on galaxy properties is deferred to future work.

  


\section{Caveats and current limitations}
\label{sect:caveats}

The emission-line maps presented in this work represent a first demonstration of the methodology developed with \texttt{GELSA} and applied to \Euclid NISP observations. The current implementation and the limited sample introduce several limitations that should be considered when interpreting the results.

\begin{itemize}

\item Current data set and redshift coverage: this work used the $0.5\,{\rm deg}^2$ \Euclid\ performance-verification observations and focused on the blue grism, for which all exposures were obtained at a single dispersion orientation. The statistical analyses were therefore restricted to $0.4< z<0.9$, where \Ha is observed with the blue grism. Future developments will extend the analysis to multiple orientations, enabling the use of the red grism and other emission lines at higher redshifts.

\item Spectral contamination: overlapping slitless spectra are not explicitly modelled in the current implementation. Although continuum subtraction with the sigma-clipped median removes much of the stellar continuum, contamination from the emission lines of neighbouring sources may remain in crowded regions.

\item Morphology versus kinematics: apparent spatial displacements in an emission-line map may arise from genuine morphological differences between the stellar and ionised-gas distributions or from velocity gradients across the galaxy. Distinguishing these effects requires observations at multiple dispersion orientations, which are available in the Euclid Deep Fields.

\end{itemize}

These limitations arise from the data analysis implementation in \texttt{GELSA} or from the specific characteristics of the observations that we analysed, rather than from intrinsic limitations of the NISP instrument or the emission-line mapping technique itself. Future developments, together with the multi-orientation observations of the Euclid Deep Fields, improved continuum modelling, and refined calibration procedures, are expected to further improve the scientific return.

\section{\label{sc:Conclusion}Conclusions}

This paper presents the first \Ha emission-line maps constructed from \Euclid blue-grism observations for galaxies at $0.4< z<0.9$. We developed a data analysis pipeline with \texttt{GELSA} to recover spatially resolved emission-line information from NISP slitless spectroscopic data. Our main results are as follows.

\begin{itemize}

\item Source-injection simulations using four blue-grism exposures yield a mean \Ha surface-brightness sensitivity of $(2.0\pm0.2)\times10^{-16}\,{\rm erg\,s^{-1}\,cm^{-2}\,arcsec^{-2}}$ at ${\rm S/N}=7$. For sources with $\YE>19.9$ above this limit, the median \Ha flux residual remains within $5\%$ of zero, with a scaled MAD of $14\%$ and $24\%$ for intrinsic FWHM values of \ang{;;0.1} and \ang{;;0.5}, respectively. The simulations also validate the recovery of radial profiles and the [\ion{N}{ii}]$\lambda6584$/\Ha ratio through 2D deblending.

\item The reconstructed maps have a median encircled-energy radius of ${\rm EE50}=0\farcs265$, corresponding to a physical scale of approximately $1.4$--$2.3\,\mathrm{kpc}$ across the redshift range over which \Ha is accessible to the NISP blue and red grisms. This spatial information enables measurements of radial \Ha profiles, identification of star-forming clumps, detection of projected velocity gradients, and assessment of spectroscopic redshift measurements.

\item Direct modelling of the \Ha--[\ion{N}{ii}] complex in the 2D maps provides a complementary measurement of the individual line fluxes. Requiring ${\rm S/N}>7$ for both \Ha and [\ion{N}{ii}]$\lambda6584$, we selected 885 galaxies and recovered a mass--metallicity relation consistent with that obtained from the 1D SPE measurements.

\item As a proof of concept, we stacked the emission-line maps of 2634 galaxies in four bins of \YE\ size and recovered \Ha\ radial profiles without requiring individual high-S/N detections. After matching the \YE\ images to the broader \Ha\ PSF, the \Ha\ profiles broadly follow the stellar continuum but remain modestly more extended, with $R_{50,\Ha}/R_{50,\YE}=1.27$, $1.12$, $1.11$, and $1.10$ from the smallest to the largest size bin. A detailed analysis of intrinsic sizes and their physical interpretation will be presented in forthcoming papers.

\end{itemize}

These results establish emission-line mapping as a complementary approach to conventional 1D spectral extraction, retaining spatial information that is otherwise lost from slitless spectra. The methodology also offers the possibility of identifying strong emission lines in median-filtered maps without prior redshift information, potentially enabling independent spectroscopic redshift determinations for faint sources not extracted by SIR, as well as spectral measurements of individual substructures within resolved galaxies. Although we present cutouts centred on individual galaxies, the methodology can also be applied to larger fields to produce continuum-subtracted emission-line maps. The present analysis demonstrates the recovery of spatially resolved emission on sub-arcsecond to arcsecond scales, without establishing an upper limit on the angular extent of recoverable sources. Our future work will extend the methodology to the red grism, multiple dispersion orientations, and additional emission lines, including \Hb, [\ion{O}{iii}]$\lambda\lambda4959,5008$, [\ion{O}{ii}]$\lambda\lambda3727,3729$, and Pa$\beta$, as well as applications to more extended systems.

\begin{acknowledgements}
LG acknowledges support from STFC and from the Henry Skynner Fellowship at Balliol College, Oxford. \texttt{GELSA} is developed within the ELSA project with grant agreement number 101135203 funded by the European Union. UK participation is funded through the UK Horizon guarantee scheme under Innovate UK grant 10093177.
  \AckEC
  \AckDRone 
\end{acknowledgements}

\bibliography{Euclid, Others_fix, DR1}
\begin{appendix}
\end{appendix}


%
%



\label{LastPage}
\end{document}